\documentclass{emulateapj}
\usepackage{subeqn}

\begin{document}


\title{Optical Cross Correlation Filters: An Economical Approach for Identifying SNe Ia and Estimating their Redshifts}


\author{Daniel M. Scolnic$^1$, Adam G. Riess$^1$, Mark E. Huber$^1$}
\affil{$^1$Department of Physics and Astronomy, Johns Hopkins University, MD 21218}
\author{Armin Rest$^2$, Christoper W. Stubbs$^2$, John L. Tonry$^3$}
\affil{$^2$Harvard-Smithsonian Center for Astrophysics, Cambridge, MA 02138}  \affil{$^3$Institute for Astronomy, University of Hawaii, Honolulu, HI 96822}


\begin{abstract}

Large photometric surveys of transient phenomena, such as Pan-STARRS and LSST, will locate thousands to millions of type Ia supernova candidates per year, a rate prohibitive for acquiring spectroscopy to determine each candidate's type and redshift.  In response, we have developed an economical approach to identifying SNe Ia and their redshifts using an uncommon type of optical filter which has multiple, discontinuous passbands on a single substrate.  Observation of a supernova through a specially designed pair of these `cross-correlation filters' measures the approximate amplitude and phase of the cross-correlation between the spectrum and a SN Ia template, a quantity typically used to determine the redshift and type of a high-redshift SN Ia.  Simulating the use of these filters, we obtain a sample of SNe Ia which is $\sim98\%$ pure with individual redshifts measured to $\sigma_z=0.01$ precision.  The advantages of this approach over standard broadband photometric methods are that it is insensitive to reddening, independent of the color data used for subsequent distance determinations which reduces selection or interpretation bias, and because it makes use of the spectral features its reliability is greater.  A  great advantage over long-slit spectroscopy comes from increased throughput, enhanced multiplexing and reduced set-up time resulting in a net gain in speed of up to $\sim30$ times.  This approach is also insensitive to host galaxy contamination.  Prototype filters were built and successfully used on Magellan with LDSS-3 to characterize three SNLS candidates.  We discuss how these filters can provide critical information for the upcoming photometric supernova surveys.

\end{abstract}





\section{Introduction}

Observations of high redshift Type Ia Supernova (Riess et al. 1998; Perlmutter et al. 1999) reveal evidence of an accelerated rate of cosmic expansion due to an apparently repulsive-like gravity.  Understanding the phenomenon of this `dark energy' provides a fundamental challenge to cosmology and
new data are vital to this endeavor.  Initial efforts have focused on 
measuring the equation of state parameter of dark energy, $w\equiv P / \rho c^2$,  
where $P$ is its pressure, and $\rho$ is its energy density.  
To determine whether the dark energy is a static, cosmological constant ($w(z)=-1$) to higher precision than past supernova surveys, future ones like the Panoramic Survey Telescope $\&$ Rapid Response System (Pan-STARRS) and the Large Synoptic Survey Telescope (LSST) will attempt to populate the SN Ia Hubble diagram with $10^4-10^6$ SNe Ia at redshifts $0.2 < z < 0.8$.  SNe Ia remain one of the best tools for understanding the properties of dark energy in the local volume because they can be discovered in large sample sizes and their individual measurement precision is high.  

Unfortunately, obtaining a SN spectrum to determine its type and redshift is observationally demanding and limits the rate at which SNe Ia can be collected.\footnote{The integration time required scales as $(1+z)^{\alpha}$, with $\alpha$ between $4$ and $6$ depending on the redshift and instrumentation. } If Pan-STARRS obtains as much time for spectroscopic follow-up as the SuperNova Legacy Survey (SNLS, Astier et al. 2006), it will still acquire spectra for only $\sim5\%$ of the sample.  Purely photometric determination of redshifts from multi-band light curves of SNe Ia (Gong, Cooray and Chen 2009), explicitly varying reddening, age and light curve shape and assuming the SN is type Ia, yields $\sigma_z \approx 0.1$ for $0.2 < z < 0.5$ and $\sigma_z$ as high as $\sim1$ for $z>0.5$.  The cosmology fitting with only this photometric data leads to a factor of 4 degradation in the dark energy figure of merit compared with the case of fitting with spectroscopic data.  While the photometric approach can be successful in distinguishing core collapse SNe from SNe Ia (Sullivan et al. 2006), the ultimate accuracy of photometric typing cannot be determined until there is a more complete characterization of the variations in core collapse SN light curves, specifically how often they resemble those of SNe Ia.  An additional complication arises when using photometric measurements to identify SNe Ia (either for follow-up or for inclusion in the Hubble diagram); they cannot subsequently be employed for the determination of cosmological parameters without introducing a selection bias.  Rather, it is important to make use of the wavelengths of spectral features (which are far narrower than broadbands) for type and redshift determination, particularly since SN classifcation is based on spectroscopy (e.g., see Filippenko 1997).

Here we propose a method of observation that will exploit the speed of photometric observations and most of the accuracy of spectroscopic observations to determine the types and redshifts of SNe Ia by using customized optical filters, each with multiple, discontinuous passbands on a single substrate.    This approach to characterizing SNe Ia is more economical and can be used to more fully harvest the yield expected from large-scale SN surveys.  In $\S 2$, we quantify the fortuitous periodicity in the spectral features of SNe Ia that enables one to distinguish them from
other common transients and determine their redshifts when imaged
with a semi-periodic passband.  SN photometry measured with two of these filters approximates the cross correlation of a coarse SN Ia spectrum with a template SN Ia spectrum commonly used to determine the type and redshift of a SN Ia candidate.  In $\S 3$, we describe our design of these filters and their use.  We discuss the advantages of these filters as a new tool for the follow-up of large-scale supernovae surveys in $\S 4$.  
In $\S 5$ we present a demonstration of the SNACC filter prototype design.

\section{Cross Correlation Formalism}
\subsection{Spectroscopic Characteristics}

Among common transients (see Fig. 1), SNe Ia are distinguished in the range $3500~-~5500~\AA$ by large undulations in flux which are deep, broad and surprisingly sinusoidal.   The origin of these spectral features is the blending of overlapping P-Cygni profiles of  singly ionized, intermediate-mass elements  (e.g., see Filippenko 1997, Baron et al. 1996).  SNe Ia are classified spectroscopically by the absence of hydrogen in their spectra and the presence of Si II (Ca II, S II, O I, Mg II  as well as Fe II are also present).  Individual spectral lines (absorption and emission) have widths of $\sim 100-150~\AA$, due to expansion velocities $\sim 10000$ km s$^{-1}$.  In contrast, the spectra of SNe II are continuum-dominated with only a few individual hydrogen P-Cygni profiles because their hydrogen shells are intact. The blended spectral features in SN Ia have equivalent widths that are typically 3-4 times greater than those in SNe II spectra at $3000-5500 \AA$. 

 Although the spectral classification of SNe Ia was originally based on the presence or absence of specific ions, 
 the loss of some features outside of the observable wavelength range at high redshifts and the typically low signal-to-noise ratios (SNRs) of such spectra preclude relying on any single ion.  As a result, the entire spectrum is considered.  Supernova type is determined by  a strong resemblance to a template spectrum $t_i(\lambda)$ at a trial redshift, $z_s$, where each type is given by a different $i$.   Following Blondin and Tonry (2007), one determines the most likely template and redshift to be those which maximize the cross-correlation function of the template and spectrum:
\begin{equation}
c(z_s,i)=\int s(\lambda) \star t_i(\lambda \times [1+z_s]) d \lambda.  
\end{equation}

Ideally, high SNR spectra would be obtained for all SNe in a survey.  In practice, as seen in the ESSENCE survey (Matheson et al. 2005) or SNLS survey (Howell et al. 2005, Balland et al. 2009), the need to obtain many spectra with limited observing time results in the collection of many spectra with low SNR degrading the information content of the measured cross correlation function to the point where multiple $z_s$ are possible.   The cross correlation function of both a high and low SNR SN Ia spectrum with SNe Ia and SNe IIP spectral templates is shown in Fig. \ref{fig:FigFourB}.  As expected, the SN Ia templates at the right redshift have a higher correlation function amplitude than the SN II templates.  For high SNR ($\sim50$ per $\sim 10.0 ~\AA$ bin), the correct redshift is found with $\sigma_z=0.01$.  However, for spectra with low SNR ($\sim 10$ per $\sim 10.0 ~\AA$ bin), a periodicity in the SN Ia correlation as a function of trial redshift is apparent, favoring a discrete set of preferred redshifts, $z_{True}$, and $z_{True} \pm n \times 0.2$ ($n=\pm1,~\pm2,\ldots$), where each local peak yields $\sigma_z=0.01$.  With a photometric redshift of $\sigma_z<0.08$, or a host spectroscopic redshift, this degeneracy can be removed.  

We can readily see the cause of this redshift degeneracy by viewing the spectra of SN Ia and SN II spectra in Fourier space.  $S(k)$ and $T(k)$, the discrete Fourier transforms of the supernova and template spectra, can be written as,
\begin{equation}
S,T(k)=\sum_{n=0}^{N-1} s,t(n)e^{-2\pi i n k /N},
\end{equation}
where we have assumed spectra are observed with N equal-sized wavelength bins with $s(\lambda) \equiv s(n)$ and $t(\lambda) \equiv t(n)$.  $n$ represents the wavelength bin such that $\lambda = n \times \frac{\lambda_{Max}-\lambda_{Min}}{N}$,  and $k=2\pi/n$.  Following Blondin and Tonry, we remove the color information in the spectra by dividing it by a cubic spline before taking its Fourier transform.    The resultant SN Ia and SN II spectra in Fourier space are shown in the top panel of Fig. \ref{fig:FigFourIa}.  The significant feature of the SN Ia spectrum is the high power at wavenumbers at $k = 6$, corresponding to wavelengths $\Delta \lambda \sim 500~\AA$, or spectral feature widths of $\sim 250~\AA$.  In the bottom panel of Fig. \ref{fig:FigFourIa}, a sine function with this wavelength is compared to a SN Ia spectrum showing their similarity in the range $2500-5500~\AA$.  This periodicity in the spectral features appears to be a fortuitous characteristic of SNe Ia.  While it is possible that multiple spectral lines that occur in a narrow wavelength range could produce such a sinusoidal shape, randomly populating the phase space with a comparable density of lines shows that a Fourier transform correlation amplitude as strong or stronger than that seen in Fig. \ref{fig:FigFourIa} occurs $<5\%$ of the time.  The distinctive shape of SN Ia spectra is recognized as `wiggly' by high-redshift SN spectroscopists even without certainty of its redshift.

The correlation amplitude for lower wavenumbers ($k < 5$) depends on the intrinsic colors of the SN and reddening.  While such spectral color information is not utilized here it may be used elsewhere for distance determinations.\footnote{In practice, SN colors are more accurately measured from photometry from which host light can be subtracted.  This also avoids chromatic slit losses which occur from long slit spectroscopy obtained off the parallactic angle (Filippenko 1982).}  The correlation amplitudes for higher wavenumbers ($k>10$) show features that are not detected with significance in low SNR spectra.
Therefore, from spectra with low but not atypical SNRs,  the only information available to determine the type and redshift of a candidate SN Ia is the power and phase of the $k=6$ mode.
The corresponding wavelength scale ($500~\AA$) for this wavenumber scales with redshift as $500~\AA \times (1+z)$.  Thus, the peak wavenumber depends on $z_s$ such that $k_p=6/(1+z_s)$, but changes little over a modest fractional change in $(1+z_s)$, as shown for $0.3<z<0.7$ in Fig. \ref{fig:FigFourAll}.  As expected, there is a large correlation amplitude near $k=4$, or $500~\AA \times (1+ 0.5 ) \approx 750~\AA$, throughout this redshift range.  Thus, a SN Ia feature centered at a true rest-frame wavelength $\lambda_{T}$, and observed at $\lambda_{O}$ due to the redshift $z_{T}$, can be confused with a similar rest-frame feature at $\lambda_{F}$ such that 
\begin{subequations}
\begin{eqnarray}
\lambda_{O}=\lambda_{T} (1+z_{T}) = \lambda_{F} (1+z_{T}+n\Delta z), \\
\textrm{where}~~~~~\lambda_{F}-\lambda_{T}=n \times 500~\AA,   
\end{eqnarray}
\end{subequations}
then
\begin{equation}
\Delta z = \frac{n \times 500~\AA \times (1+z_{T})}{\lambda_{T}}.
\end{equation}

For example values of $\lambda_T=3600~\AA$, $z_T=0.5$, then $\Delta z \approx \pm 0.2$ for $n=\pm1$.  Therefore, the undulations in SN Ia flux with periodicity of $500~\AA \times (1+z)$ cause the redshift degeneracy of $\Delta z \sim 0.2$, as in Fig. \ref{fig:FigFourB}.  

SNe II spectra do not share this periodicity and their normalized correlation amplitude at $k=6$ is 3.5 times lower than that of a SN Ia, as seen in Fig. \ref{fig:FigFourIa}.  Thus, measuring the power of SNe Ia and SNe II spectra at $k=6$ allows one to distinguish between these two types of SNe and the phase of that power provides the redshift of a SN Ia.  

In practice, the determination of SN Ia type and redshift may be acquired with just two samplings of the cross-correlation function at different trial lags.  An example of this is shown in the right panel of Fig. \ref{fig:FigFourB}.  Two samplings break the degeneracy of the phase, $\pi$, of the sine curve and the two measurements are optimally made with phases $\phi_1=0,\phi_2=\pi/2$ ($ \Delta z \approx 0.05$ in redshift space).  Since these two measurements are out of phase, at least one of the amplitudes measured reveals high power at $k=6$ if the SN type is Ia.  If a redshift prior (photometric or host spectroscopic) with $\sigma_z <0.1$ is also available, one can also confidently determine a SN Ia redshift with $\sigma_z\approx0.01$, similar to the precision of a local peak in the SN Ia cross-correlation function.

\subsection{Imaging Approach}
The convolution theorem states that the convolution of an input spectrum and a template is equivalent to the product of their Fourier transforms:
\begin{equation}
F( s(n) \ast t(n))=b \times S(k) \times T(k).
\end{equation}
Our definition of the cross-correlation, given in Eq. 1, is similar to a convolution and thus Eq. 5 is applicable.  Since low SNR spectra have most of their information in the power and phase of the $k=6/(1+z)$ mode, the template that maximizes the product of the fourier transforms, and thus the amplitude of the cross-correlation, should have the majority of its power in the $k=6/(1+z)$ mode as well.  Therefore, we can replace the SN Ia template spectra with sinusoidal functions with $\lambda =500\times(1+z)$ with no loss in information in the cross-correlation function.

These simplified templates, $t(\lambda, \phi_1)$ and $t(\lambda, \phi_2)$, can now be replaced with optical transmission functions (i.e., optical filters) which can be used to evaluate the cross-correlation amplitude and phase by $\textit{direct imaging}$.  We now define $CC$ as the SN magnitude through the sinusoidal filter and $B$ as the SN magnitude through a broadband filter.  $CC-B$ then measures the cross-correlation amplitude as shown in Fig. \ref{fig:FigFourB} and can be expressed as
\begin{equation}
CC-B-\mbox{ZP}=-2.5 \times \left ( \log_{10} \left (\frac{F_{CC}}{F_{B}} \right) \right)-\mbox{ZP}
\end{equation}
where $F_{CC}$ is the flux through the template filter, $F_{B}$ is the flux through the broadband filter and $ZP$ is a zeropoint that will be discussed later.

  Here after we will refer to optical filters which mimic $t(\lambda, \phi)$ as SuperNovAe Cross Correlation (SNACC) filters.  With two such filters, these measurements identify a SN in the $CC_1-B$ (mag) versus $CC_2-B$ (mag) plane.  If one of these SNACC filters is chosen to be $\pi/2$ out of phase with another filter, then SNe Ia, over a moderate redshift range, delineate a circle in this correlation space with values of $CC-B \sim 0.25$ mag when the spectral features of the SNe are correlated with the template bandpasses, and $CC-B\sim-0.25$ mag when the features are anti-correlated with the template bandpasses.  The period of the SN Ia circle with radius $0.25$ mag is the same as the degeneracy of the cross-correlation versus redshift relation for low SNR spectra: $\Delta z=0.2$.  With a host photometric redshift ($\sigma_z<0.08$), two measurements with $\sigma_{CC-B} \sim 0.04$ mag in $CC-B$ space reveal the redshift of a SN Ia to $\sigma_z=0.01$, the precision inherent in the SN Ia correlation curve.  

	All SNe II (IIP, IIL, IIn) remain clustered near $CC-B \sim 0$ which provides the means to distinguish them from SN Ia.  Since SNe II have low correlation amplitudes at $k=6/(1+z)$ for $0<z<1$ in the observed frame for a broad range of $z$, there is little difference between the relative flux in a template filter and the broadband filter.  In addition, the few features of SNe II have lower equivalent width with respect to the mean flux than those of SNe Ia, so that if one or two features (at most) align with the template filter teeth, the value $|CC-B|$ is still smaller for SNe II than for SNe Ia.  

\section{Implementation}
Optical filters with multiple passbands may be designed to mimic the spectral features of SNe Ia and used to constrain the type and redshift of SNe.  While important properties of these filters have already been discussed (e.g. separation between the passbands $\sim 500~\AA \times (1+z_{Median})$ ), the precise determination of the filter specifications requires greater rigor.  Here the optimal filters are determined as those which maximize the precision of the determination of redshift and SN type.

We define an extended broadband filter, $b'$, to transmit the sum of flux from Sloan g' and r' so $F_{b'}=F_{g'}+F_{r'}$.    For convenience, we define $ZP\equiv CC_{G8V}-B_{G8V}$ for Eq. 6 so that a G8V star has $CC-B-ZP=0$ mag.  Since a G8V star has a thermal spectrum with $T=5310$ K, and similar colors as the spectra of SNe Ia at $z=0.5$\footnote{Supernovae have temperatures $\sim 10,000~\mbox{K}$ .  At $z=0.5$, then $10,000~\mbox{K}~/(1+z) \sim 6500~\mbox{K}$.}, setting $ZP=CC_{G8V}-B_{G8V}$ provides a useful reference for measurements at $(CC_1-B, CC_2-B)=(0,0)$.  Stars with the same spectral type but different luminosity class will differ by $<0.03$ mag.

The flux $F_{CC}$ through a filter $P$ with discontinuous passbands or `teeth' $p_1, ~p_2 \dots$ is given by
\begin{equation}
F_{\mbox{CC}}=\int ~ S(\lambda(1+z))P(p_1,p_2 \dots) d\lambda.
\end{equation} 

 A single passband, or `tooth', $p_i$, can be described with three parameters: a central position ($\lambda_i$), a  width ($\Delta \lambda_i$) and a transmission fraction ($r_i$).  The set of values $\mathbf{\lambda}, \mathbf{\Delta \lambda}, \textbf{r}$, for a total of $3\times2~\textrm{(filters)}\times n~\textrm{(passbands)}$, can be varied to optimize a merit function and thus the design of a filter.  

To construct a useful merit function we first simulate a generic SN survey with a few assumptions about the properties of SN and their redshift distributions.  For later optimizations, the survey assumptions are designed for a specific survey, e.g., Pan-STARRS-I.  Our simulated population is composed of $5000$ SNe Ia and $5000$ SNe II.  The SN Ia spectral templates from Hsiao et al. 2008 are used for our SNe Ia spectral template; the spectrum of SN IIP 1999em (Baron et al. 2000) is used as a SNe II spectral template.  SNe IIn and SNe IIL spectra have even narrower features than those of SNe IIP, and should be even easier to discriminate from SNe Ia, so SNe IIP will represent an upper limit for difficulty in discriminating SNe Ia from SNe II.  SNe Ib/Ic/Ibc, as well as other transients, are rarer and will be deferred to $\S 3.5$. The number of simulated SNe grows with the redshift due to increasing volume, then decreases with redshift as the brightnesses of all SNe no longer exceed the magnitude limit.  We find the redshift distribution of our simulated SN Ia survey can be approximated as a truncated Gaussian distribution for a limit of $r'=25$ mag and mean SN Ia brightness of $-19.16$ mag (Richardson et al. 2002): $\bar{z}=0.55,~\sigma=0.15,~z_{min}=0.30,~z_{max}=0.75$.  The SNe II redshift distribution, with mean brightnesses for SNe IIP, SNe IIL, SNe IIn of $-16.6,~-17.4,~-18.78$ mag, can be approximated as $\bar{z}=0.45,~\sigma=0.15,~z_{min}=0.20,~z_{max}=0.7$.  For each SN, we determine its $CC-B$ values from its template spectrum, redshift and the filter parameters.  We then simulate noisy measurements with our assumed photometric errors.

Because our goal is to find the type of SN, and if a Ia, its redshift, the following chi-squared statistic is minimized with respect to type and z for an individual SN:
\begin{eqnarray}
\chi^2 (\mbox{type},z)&=& \left (\frac{(CB_{1}-CB_{1z})}{\sigma_{1}} \right )^2+ \left (\frac{(CB_{2}-CB_{2z})}{\sigma_{2}} \right )^2 \nonumber \\
& &+\left (\frac{z_p-z}{\sigma_z} \right )^2
\end{eqnarray}
where $CB_1$ and $CB_2$ are simulated measurements for $CC_1-B$ and $CC_2-B$, $CB_{1z}$ and $CB_{2z}$ are the expected values of $CC-B$ positions for a SN of a particular type and redshift, $\sigma_{1,2}$ are the simulated measurement errors of $CB_1$ and $CB_2$, $z_p$ is the available redshift prior and $\sigma_z$ is its error.  The measurement errors, $\sigma_{1,2}$, include image calibration errors and photon statistics.  We estimate the errors resulting from empirically calibrating the $CC-B$ value of the supernova from stars in the field to be $\sim0.02$ mag (see \S 5). The photon statistics are calculated from the brightness of the SN and the amount of time alloted for imaging on the Low Dispersion Survey Spectrograph (LDSS-3, upgraded from LDSS-2, Allington-Smith et al. 1994), on the 6.5-m Magellan Clay Telescope.  

We also expect some variation of the SNe Ia spectra, and hence CC-B, which correlate with the luminosity-light curve shape parameter or `stretch', the age of the SN and with individual reddening (Guy et al. 2007).  The remaining, intrinsic variation of SNe Ia spectra is small compared to variations arising from these three parameters (Fig. 4 of Guy et al 2007).  

Depending on the circumstances in which the SNACC filter data will be analyzed, we may be able to make use of knowledge about the age, stretch and reddening of a SN Ia.   The available SN light curve may be complete, sparsely-sampled, or consist of only the discovery point.  If a well-sampled light curve is obtained, then a SN Ia template with the corresponding age, stretch and reddening of the SN candidate would be used to improve the chi-squared diagnostic (Eq. 8).  Likewise, $\sigma_z$ depends on whether a host galaxy redshift is known photometrically (e.g., $\sigma_z=0.08$), spectroscopically ($\sigma_z=0.01$), or unavailable ($\sigma_z= \infty$).  Classification resulting for these various follow-up scenarios as well as the effects of age, stretch and reddening on the success of determining the type and redshift will be discussed in $\S 3.3$, $\S 3.4$.

\subsection{Merit Metrics for Optimizing Filters}

The two categories of metrics to be used for optimization are the success of determining the type of SN and if Ia, its redshift.  Let's assume the two most common SN populations, SNe Ia and SNe II.  We define $Y_{Ia}$ as the fraction of known SNe Ia that are inconsistent with SNe II.  Similarly $Y_{II}$ is the fraction of SNe II that are inconsistent with SNe Ia.  The \textit{purity} of the SNe Ia sample that would be added to the cosmological sample is then: 
\begin{equation}
\mbox{Purity}=\frac{Y_{Ia} \times N_{Ia}}{Y_{Ia} \times N_{Ia} + (1-Y_{II}) \times N_{II}}
\end{equation}
where $N_{Ia}$ and $N_{II}$ are the numbers of known SNe Ia and known SNe II respectively.

The success in determining the redshift for SNe Ia is quantified by $\sigma_{z_{spec}-z_{SNACC}}$, where $z_{spec}$ is the known redshift and $z_{SNACC}$ is the redshift that minimizes the chi-squared diagnostic (Eq. 8).

\subsection{Prototype 3-Tooth Design}

To design prototype filters, we assumed the availability of a photometric redshift prior with $\sigma_z\approx0.08$ at $0.3<z<0.8$ and a light curve based age of the SN (which is $\pm5$ rest-frame days from peak brightness when the $CC-B$ mags are obtained).  We also assumed integration times would be sufficient to reach combined photon statistics and calibration errors of $\sim\sigma_{1,2}\le0.05$ mag and that the photon statistics would depend on avoidance of the skylines in the passbands.  

We first attempted to design a simple filter model with just two discontinuous passbands.  With filters of this kind, the ability to discriminate type of SN is not strong over a redshift range of $\Delta z=0.5$.  Next, we searched for a three tooth design for our SNACC filters, which is the solution we show here.  A filter with four teeth, designed by optimizing both the throughput of the filters and the discrimination power, will be discussed in $\S 6$.   

A pair of 3-tooth filters, which maximizes the purity of SNe Ia, is shown in Fig. \ref{fig:FigFilt3} and the expected positions of SNe Ia and SNe II are shown in Fig. \ref{fig:FigCirc3}.  The widths of each tooth is $\sim160~\AA$, but with different transmission heights ($0.42,~1.0,~0.66$), thus these 3-tooth SNACC filters have an effective width of $\sim 325 \AA$, or $\sim33\%$ that of a broadband filter.\footnote{Elsewhere (Scolnic et al. 2010) we present a 4-tooth design which a includes a 4th tooth between 1st and 2nd as seen in Fig. \ref{fig:FigFilt3} and achieves a width of $600~\AA$ without loss in purity.}  The mean wavelength 
\begin{equation}
<\lambda>=\frac{\int \lambda T(\lambda) d\lambda}{\int \lambda d\lambda} 
\end{equation}
of the two SNACC filters ($<\lambda>=5750,~5650~\AA$ respectively) and the broadband filter ($<\lambda>=5550~\AA$) are quite similar which greatly limits the effect of reddening on the observed $CC-B$ values.  Also, the bandpasses of the second filter avoid all major skylines, but because the center bandpass of the first filter is near the $5887~\AA$ Na D line, this filter transmits $20\%$ more sky flux than the first.
As shown in Fig. \ref{fig:FigFilt3}, SNe Ia $CC-B$ mags trace roughly concentric circles tracking counter-clockwise with increasing redshift.  As expected, SNe II lie closer to the origin due to their lack of spectral features as strong as SNe Ia, with an average separation from the origin of $\pm 0.05$ mag.  For measurement errors of $\sigma_{m}=0.04$ mag (as we obtained with the first use of the SNACC filters, discussed in \S5.2), the SNe Ia and SNe II are separated in $CC-B$ space by $\sim3.0-5.0 \times \sigma_m$.  The actual fraction of SNe Ia correctly classified, $Y_{Ia}$, is found to be $99.0\%$ and the purity of SNe Ia, assuming equal sizes of SNe Ia and SNe II populations, is $98.0\%$.  Measurement errors of 0.03 and 0.05 mag, which we consider lower and upper limits for our measurements, result in purities of $\sim99.0\%$ and $\sim95.0\%$ respectively.

SNe IIL (using spectral templates that are updated versions of those given in Nugent et al. 2002\footnote{ \texttt{http://supernova.lbl.gov/$\sim$nugent/nugent\_templates.html}}), like SNe IIP, have an average distance from the origin of $CC-B$ space of $\approx0.05$ mag.  SNe IIn spectra are mostly continuum, with few narrow lines that form in low velocity circumstellar matter.  Since the width of these lines is so small, the average distance of SNe IIn in $CC-B$ space from the origin is $\approx 0.01$ mag.

The histogram of $\Delta z =z_{spec}-z_{snacc}$ for all the SNe Ia is shown in Fig. \ref{fig:FigHist}, assuming a host photometric error $\sigma_z=0.08$ and measurement errors $\sigma_{1,2}=0.04$ (mag).  The redshift distributions are centered around $\Delta z=0$ with a standard deviation around the center of $\sigma_z=0.01$.  Due to the cross-correlation redshift degeneracy and the limited assumed precision of the host photometric redshift, there are over-populated tails at $\Delta z=\pm 0.2$ that hold $\approx 3\%$ of the distribution.  Without a photometric redshift, the tail population is $\approx 45\%$.  An error in $z=\pm0.2$ should be easily distinguishable in the Hubble diagram since the difference in distance for two SNe Ia with this redshift separation is $\sim 1.0$ mag.\footnote{$\Delta d = 5 \log \frac{cz_T}{H_0}-5 \log \frac{c (z_{T}+0.2)}{H_0}= 5\log \frac{z_T}{z_{T+0.2}} \approx 1.0~\textrm{mag for }z_T\sim0.5$.}

\subsection{Effects of Light Curve Shape, Age and Reddening on $CC-B$ magnitudes}

Variations in SN Ia spectra are mainly due to variations in reddening, light curve shape and age.  The effects of ignorance of these parameters (using the SNe Ia variations modeled in Guy et al. 2007) on our purity and redshift errors are shown in Fig. \ref{fig:FigFactors} and Fig. \ref{fig:FigFactorsR}, respectively.  Typical usage of SNACC filters would avoid these errors using light curves to choose a SN Ia spectral template corresponding to the appropriate age, light curve shape and reddening.

Due to the near-equality of the mean wavelength of the SNACC filters and the broadband filters, the fraction of SN Ia correctly classified at most redshifts is quite insensitive to reddening, affecting $Y_{Ia}$ at the $1\%$ level, as shown in Fig. \ref{fig:FigFactors}.  Selection purity is also insensitive to the light curve - luminosity relation, varying by $\sim2\%$ (except for $z>0.75$).  Ignorance of the SN age has a stronger impact on $Y_{Ia}$ and limits the time range in which it is optimal to use SNACC filters for follow-up.  The purity is very robust for ages of $-7$ to $+5$ days but degrades for older SNe Ia.  $Y_{Ia}$ is $<90\%$ for SNe with $z\le0.35$ past $+5$ days and for SNe with $z\ge0.75$ past $+8$ days.  For SNe Ia with $0.35<z<0.75$, $Y_{Ia}$ on average is $\sim95\%$ up to an age of 10 days, and even for ages past this, $Y_{Ia}$ for parts of the redshift range remains as high.  Prior to 5 days before peak, the overall purity remains $>95\%$, but SNe Ia with $z\sim0.55$ (for $-10$ days) or $z\sim0.65$ (for $-15$ days) will approach the origin in $CC-B$ space.  While it is difficult to simulate how peculiar SNe Ia with similar early ages and high redshifts appear in $CC-B$ space because of the limited set of UV spectra, SNe like SN 1991T (Jeffrey et al. 1992), for redshifts $0.25<z<0.40$, are located in $CC-B$ space with the same phase and at most a $\sim0.05$ mag relative distance to the origin as their normal SNe Ia counterparts (and $Y_{Ia}$/$\sigma_z$ values of $92\% / 0.012$).  For these peculiar SNe Ia, the strengths of the spectral features vary from those of normal SNe Ia, but since three separate passbands are correlated with the spectrum, the overall difference between these two categories of SNe Ia in $CC-B$ space is not large.  The effects of age, light curve shape, and reddening are most noticeable at the limits of our redshift range ($z\sim0.35,0.75$) since the spectral features at these redshifts that correlate with the passbands are weak compared to the correlating features at $z\sim0.5$.

Both reddening ($0.0  <E(B-V) < 0.4$)  and stretch ($0.85 < s < 1.15$) cause errors in $| z_{SNACC}-z_{Spec} |$ that are all $< 0.02$ as shown in Fig. \ref{fig:FigFactorsR} .  Anytime before 5 days past peak, age has a limited effect on the redshift errors ($<\sim0.02$), but afterwards the effect is much more significant (typical errors $\sim 0.05$) for all redshifts.

\subsection{Usage Scenarios}

SNACC filter measurements may be used in conjunction with other data to yield more precise results.  For example, the host galaxy of the SN may be apparent, and the multi-band light curves may have been acquired.  If a host galaxy is visible, a photometric redshift ($\sigma_z=0.08$) or spectroscopic redshift ($\sigma_z=0.01$) may be obtained.  If a light curve has been acquired, values of the age, stretch, reddening and photometric redshift ($\sigma_z/(1+z)=0.06$, based on Sullivan et al. 2006) can be fit on the condition the SN is Ia.  An available photometric or spectroscopic redshift breaks the $\Delta z = 0.2$ cross-correlation degeneracy of SNe Ia and knowledge of the age, stretch and reddening reduces the spectral variations because the appropriate spectral template can be used.

As realistic examples, the Medium Deep Pan-STARRS survey, which images the same area of sky every four days, should provide well-sampled light curves. In contrast, the 3$\pi$ Pan-STARRS survey covers the entire sky only twice a month, and will not provide light curves well-sampled enough to fit accurate values of the type, stretch, reddening or age of the SN.

The combinations of possible redshift priors (photometric, spectroscopic, none) and light curve based priors (yes, no) make up six distinct scenarios.  Table I presents these scenarios and the results expected after SNACC follow-up.  We simulate a realistic set of SNe Ia with a uniform distribution of ages within $\pm 5$ rest frame days from peak, a gaussian distribution of stretches with $s=1$ and $\sigma_s=0.1$ and an exponential distribution of reddening values, with a peak at $E(B-V)=0$ and $\tau=0.1$.

Table I shows that the fraction of SNe Ia correctly classified and the redshift determination both improve with the amount of information one obtains about the supernova or host galaxy.  With a well-sampled light curve and host galaxy but no SNACC filter measurements, we assume a baseline of $Y_{Ia}=80\%$ and $\sigma_z/(1+z)\approx0.06$.  $Y_{Ia}$ is $97.0\%$ if one obtains a well sampled light curve of the SN because then one can use the correct template in the chi-squared diagnostic of Eq. 8.  Similarly, the precision of the redshift determination is also greater if a well sampled light curve is obtained.    The best case is the scenario in which one obtains a spectroscopic redshift from the host galaxy in addition to a well sampled light curve of the SN.  Here, $Y_{Ia}$ is $99.0\%$, although this is not significantly better than the case in which one obtains only a photometric redshift of the host.  Generally, if one gathers supplementary information
about the SN in addition to the SNACC filter measurements, one achieves $Y_{Ia}>\sim95\%$ with $\sigma_z<0.015$.

\subsection{Other Transient Candidates} 

Until now, only the properties of SNe Ia and SNe II in $CC-B$ space have been considered because they are the most common and therefore their separation is most important to good SN Ia selection.  There are other types of transients, and we list here some that may be confused with SNe Ia (Matheson et al., 2005) and their characteristics in $CC-B$ space.

\begin{description}
\item[SNe Ib/Ic/Ibc] The progenitors of SNe Ib/Ic/Ibc (hereafter called SNe Ib-c) are believed to be stars stripped of their original H-rich envelope (Georgy et al. 2009) and the spectral features of these supernova are similar to those of SNe Ia (as seen in Fig. 1).  The SNe Ib-c spectral templates we use are updated versions of those given in Nugent et al. 2002.  The characteristic spectroscopic difference between Type Ia and Ib-c SNe is the deep absorption trough at $6150~\AA$ in Type Ia spectra, which is due to the blue-shifted Si II $\lambda \lambda6347,6371$ feature (Homeier 2005).  Unfortunately, this is redshifted out of the optical passband for $z>0.4$, making it difficult to distinguish SNe Ia from SNe Ib-c.  Although there are other spectroscopic differences between SNe Ia and Ib-c (e.g. Si II $\lambda4130$, Coil et al. 2000), they are too slight for the two types with the same redshift to appear in different locations in $CC-B$ space.  There are two branches of redshifts, $0.1<z<0.18$ and $0.30<z<0.38$ in which the SNe Ib-c are clustered in the center, but for other redshifts, their $CC-B$ positions are often less than $0.05$ mag away from a SN Ia with the same redshift.  However, they are rare enough ($<3\%$ of photometric SN Ia candidates targeted for spectroscopy in the ESSENCE project (Matheson et al. 2005)) to not simultaneously affect our SN Ia purity.  We also note that SNACC filter observations of SNe Ib-c fair no worse than typical high-redshift spectroscopy whose SNR is frequently (Riess et al. 1998, Matheson et al. 2005) too low to identify Si $\lambda 4130$ and which is not even present for many SNe Ia.  For the set of ESO/VLT spectroscopy of SNLS candidates (Balland et al. 2009), 38 out of 124 of the classified SN Ia spectra were called `probable'  SNe Ia because other types, in particular SNe Ic, could not be excluded given the SNR or the phase of the spectrum.  In any case, one learns the correct redshift for SNe Ib-c which aids their elimination from the Hubble diagram with Bayesian techniques (Kunz et al. 2006).
\item[AGN] There are several subclasses of AGN (e.g. Seyfert 1, Seyfert 2, LINER) and the AGN in each subclass have unique spectral features (Osterbrock 1989).  Here we use the set of 12 AGN spectra obtained by the ESSENCE project which are particularly relevant because they were photometrically identified as SN Ia candidates.  In the project, AGN were discovered with redshifts $0.18<z<2.6$, so we must include a large redshift range when determining the location of AGN in $CC-B$ space.  For the ESSENCE set of spectra, over the redshifts in which the spectra cover the entire optical range, the AGN will deviate from the origin $>0.05$ mag only $\sim5\%$ of the time.  Thus, $>11$ out of 12 ESSENCE AGN would have been rejected from the SN Ia sample.  For spectra at select redshifts (e.g. spectra similar to those of Seyfert 1 AGN at $z\sim2.57$, see Francis et al. 1991 for spectral templates), the position of AGN in $CC-B$ space may deviate as far from the origin as SNe Ia.  Many AGN can be discriminated from SNe Ia with broadband color measurements and possibly a time-series history, and SNACC filters would provide an additional method to differentiate the AGN from SNe Ia.
\item[Variable Stars]  During the course of the ESSENCE project, as well as other recent supernova surveys, several variable stars as well as unclassified stars were discovered.  In the spectra published from the ESSENCE project, the spectra of the stars were mostly continuum and are located near the origin of $CC-B$ space with little deviation.  Thus, rarely should any variable stars be confused with SNe Ia.

\end{description}

\section{Observations}
\subsection{SNACC Filter Fabrication for LDSS-3}
The technology used to fabricate SNACC filters, with multiple discontinuous passbands on a single substrate, has become popular for a wide range of applications in many fields, like fluorescence microscopy in biology (e.g. Yelin and Silberberg, 1999).  We canvassed multiple vendors but ultimately Iridian Spectral Technologies Ltd.\footnote{For more information see \texttt{http://www.iridian.ca/}} was best able to match our specifications for the prototype within a reasonable cost.  

Iridian constructs Thin Film Filters (TFF, see Macleod 2001, for a review) by layering materials of alternating refractive index with thicknesses on the order of visual wavelengths, $\sim5000~\AA$, to produce optical filters that transmit and/or reflect specific wavelengths.  A thin-film coating is a stack of such layers having boundaries of greater refractive index (reflecting light $180^\circ$ out of phase) and less refraction index (reflecting light in phase) that if at precise spacings will modify the reflected and transmitted components by interference.

The filter design team at Iridian use a proprietary design and process control technology, PrecisionSpectrum@IST, to develop the optimal multilayer design for the requested filter specifications and requirements (both physical and optical). The SNACC filters were designed with a TFF layering on the front-side and simple two-layer anti-reflective type coating on the backside. 

The filters were specified for use in the Low Dispersion Survey Spectrograph (LDSS-3, upgraded from LDSS-2, Allington-Smith et al. 1994), on the 6.5-m Magellan Clay Telescope.   Our specified transmission function of the filters was adjusted to include the effect of the camera efficiency.  Our request, Iridian's expected design, and the design they achieved are shown in Fig. \ref{fig:FigIridian}.  The bandpasses of the filters Iridian produced had width, height and position values that were all within $5\%$ of our specifications.  

In the LDSS-3 setup, an input f/11 beam enters a collimator before passing through a SNACC filter. The light is then focused by the camera onto an external detector with a final focal ratio of f/2.5.  Due to off-axis rays, there is a correlation between the incident angle on the filter and the radial field position.  For 2/3 of the field, there is a maximum angle of incidence of $\sim4.6^\circ$ of the beam on the filter.  Because of the construction of the interference filter, there is a shift in the effective filter wavelength, to the blue (i.e. at $6000~\AA$, this is a shift of about 10A).  The filters were analyzed in the optics lab at Johns Hopkins University to measure their transmission functions in a similar setup with a tungsten halogen lamp, silicon photodiode, and collimator.  We measured filter transmissions which were very close to what Iridian stated they achieved in a parallel beam, and that the out of band rejection was $>99.5\%$ over the wavelength $3000-11000~\AA$. When collimated beams have an angle of incidence of up to $5^\circ$ on the filter, the shifts in the bandpasses were fairly coherent with values up to $8~\AA$, negligible for our purpose.  

\subsection{SNACC Filters Imaging and Spectroscopy with LDSS-3}

To test the utility of the filters for SN classification, we observed three supernovae and spectrophotometric field calibration stars.   Several supernova candidates were discovered in the SuperNovae Legacy Survey (SNLS, Astier et al. 2006) with the Megaprime camera at the Canada-France-Hawaii telescope after observations on 2008 May 1 (previous observation date was 2008 April 14) in the SNLS D2 field (RA 10:00:29, DEC 02:12:21, 1 square degree).  Along with the candidates, SNLS provided the candidates' photometric redshift if it was assumed they were SNe Ia using the SN's broadband color measurements, g,r,i,z measurements around peak, and photometry of the host.  A significant advantage of the SNACC filters over long-slit spectroscopy is that they easily allow for host light subtraction.  In this case, however, we chose to follow-up reasonably isolated SNe since we did not have time to obtain SNACC filter host templates.  We chose the candidates 08D2nu (NU) and 08D2oe (OE) because their available (though minimally sampled) light curves exhibited the pre-peak rise and appropriate $g-r$ colors of SNe Ia light curves (sensitive to reddening), and 08D2jh because its light curve exhibited the post-peak plateau of a SN IIP light curve.  Referring back to \S 3.4, the follow-up scenario in which our observations of these candidates took place is the one in which a light-curve is sparsely populated and there is a photometric redshift from the host galaxy.  The calibration stars, a G8V star ($r'=16.7$ mag) and a M4V star ($r'=17.4$ mag), were selected from the Sloan Digital Sky Survey Data Release 7 (SDSS-DR7) (Abazajian et al., 2009), which provided photometry and spectra.  The stars' types were determined from cross-correlating their spectra with spectra from the Gunn~\&~Stryker spectrophotometric atlases (Gunn~\&~Stryker, 1983).  The G8V star was chosen because its $CC-B$ values should be near the origin, and the M4V star, with its strong spectral features, was chosen because it should deviate as far from the origin as SNe Ia in $CC-B$ space and would be helpful to verify the transmission function in the beam.  As we will discuss later, the  $CC-B$ values of the G star (with $g-r<1$) only depend on its luminosity class because of the relation between luminosity class and $g-r$ color, whereas the $CC-B$ values of the M star (with $g-r>1$) depend strongly on its luminosity class because of the star's strong spectral features.   

The observations obtained for the SNe and calibration stars are given in Table 2 and 3 respectively.
We observed SN NU and SN OE on UT 2008 May 3 with the SNACC filters on LDSS-3 for 300 seconds each, and SN JH on May 5 for 300 seconds as well.  The G8V and M4V stars were observed for exposure times of 180 seconds on UT 2008 May 3.  Flat-field images and bias frames were also acquired.  A custom pipeline was used to perform photometry of the bias-subtracted, flat-fielded images, using aperture photometry via the APER routine in IDL and point-spread function photometry via the DAOPHOT routines in IDL.

We also acquired spectra of the SNe NU, JH, OE on UT 2008 May 4, 5, 9 with the VPH-All grism on LDSS-3 with a $1''$ slit for integration times of $2\times1200$ s, $2\times900$ s, and $2\times1800$ s respectively. The VPH-All grism (400 lines/mm) allowed us to acquire 4000 - 9900 $\AA $  spectra with an average resolution $\lambda / \Delta \lambda \approx 860$ across the entire chip.  Dispersion along the chip was 1.9 $\AA $ /pixel.  All spectral observations were accompanied by HeNeAr arc lamps exposures to measure dispersion and the pixel to pixel response was removed with flats from a quartz lamp.  The spectral data were reduced using the standard IRAF\footnote{IRAF is distributed by the National Optical Astronomy Observatories, which are operated by the Association of Universities for Research in Astronomy, Inc., under cooperative agreement with the National Science Foundation.} packages including APALL (extraction), DISPCOR (assigning the dispersion function from the calibration arc lamp) and CALIBRATE (using sensitivity function to flux calibrate the spectra).  Flux calibration (using STANDARD) of the spectra was performed by means of spectrophotometric standard stars observed at similar airmass on the same night as the SN.  For all the spectra observed, the slit was always aligned along the parallactic angle to avoid differential chromatic refraction.

\section{Analysis of Observations}
\subsection{SNACC Filter Calibration: Photometry for G' and M' stars}

Using spectra from the Gunn~\&~Stryker spectrophotometric atlases (Gunn~\&~Stryker, 1983) and the known transmissions of the broadband and SNACC filters, we use the relation between the $CC-B$ values and $g-r$ values for standard stars to determine their zeropoints.  

The $CC-B$ values of the stars are linearly related to their $g-r$ values, as seen by the diamonds in Fig. \ref{fig:FigGShift},\ref{fig:FigMShift} such that one can write the relation for stars
\begin{equation}
CC-B=m (g-r) + Y~~~\textrm{for}~g-r<1.0 
\end{equation}
where $CC-B$ has already been offset by the zeropoint as defined previously, $ZP\equiv CC_{G8V}-B_{G8V}$.   The slope and intercepts can be found by a least-squares fit: $(m_1,Y_1)=(-0.14, 0.04), (m_2, Y_2)=(-0.19,0.07)$.  These relations are highly linear for $g-r<1.0$ and have a dispersion of less than $0.02$ mag.  Stars with $g-r>1.0$ generally are M-type stars with strong spectral features that break the $CC-B$/$g-r$ relation.

For a set of $n$ observed stars, each with $CC$, $B$, $g$ and $r$ values ($g$ and $r$ have been corrected for zeropoint offsets), the $CC-B$ color zeropoints $\mbox{ZP}_1$ and $\mbox{ZP}_2$ for a field can be determined from:
\begin{eqnarray}
CC_1-B-\mbox{ZP}_1 =-0.14 (g-r) +0.04\\
CC_2-B-\mbox{ZP}_2 =-0.19 (g-r) +0.07
\end{eqnarray}

To find the zeropoints, one minimizes $\chi^2$ with respect to ZP
\begin{equation}
\chi^2=\sum_{i=0}^n~\frac{(\mbox{ZP}-\left((CC_i-B_i)-m(g_i-r_i)+Y\right)^2}{\sigma_i ^2}
\end{equation}
where $g_i-r_i < 1.0$ and $\sigma_i$ is the quadrature sum of star i's measurement errors.  Having determined $\mbox{ZP}_{1,2}$, the $CC-B$ values of the SNe in the field can be placed in our diagnostic $CC-B$ space.

The results of this calibration method for the G8V field and M4V field are shown in Fig. \ref{fig:FigGShift} and Fig. \ref{fig:FigMShift}.  The $ZP$ values found for the G8V field were $ZP_1=0.33\pm0.01$ and $ZP_2=0.24\pm0.005$.  The $ZP$ values found for the M4II field were $ZP_1=0.33\pm0.015$ and $ZP_2=0.22\pm0.01$.  

The $CC_1-B$ and $CC_2-B$ values of all the stars in the G8V field and M4V can be combined, as can be seen in Fig. \ref{fig:FigMArr} and Fig. \ref{fig:FigGArr}, and we can compare the values of the  G8V and M4V stars with their synthetic spectrophotometric measurements (spectra from SDSS DR7).  To determine the total photometric error, the calibration error of the stars were added in quadrature with the photon statistics error.  The G8V and M4V stars were $\sim17$ mag in the r' band, and the total photon statistics error from the SNACC filters and broadband filters was $\sim0.015$ mag.  As seen in Fig. \ref{fig:FigMArr} and Fig. \ref{fig:FigGArr}, the synthetic measurements are within one standard deviation of the photometric measurements.

\subsection{Results of Imaging and Spectroscopy}

The extracted SN spectra, as observed with the LDSS-3 spectrograph, are presented in Fig. \ref{fig:FigAllSpec}.  Bluer than $4300~\AA$, the SNR$<5$ due to the decreased sensitivity of the spectrograph and lower object flux with respect to the sky.  Template spectra of the same type and age were used to graft this part of the spectrum.  The SNID program (Blondin and Tonry, 2007) was used to determine that NU and OE are both SNe Ia, with redshifts $z=0.56\pm0.01$ and $z=0.42\pm0.01$ respectively, and JH is a SN IIP with $z=0.23\pm0.01$.  The quality of correlation is determined by the rlap quality parameter, which is the product of the correlation height-noise and the spectrum overlap parameter, and is 0.63, 0.68 and 0.74 for NU, OE and JH respectively.  Using the extracted spectra as well as transmission functions of the SNACC filters and broadband filters,  the synthetic spectrophotometric $CC-B$ values of the SNe can be evaluated and compared to the observed values.

Details of the SNe ages, broadband colors, and redshifts (host galaxy photo-z, SN photo-z from broadband colors, and spectroscopic) are given in Table 4.  Observations of the SNe were sky-noise limited, so the photon statistics error is determined from the dispersion of the photometry when a model psf with the source's amplitude is placed at random locations in the field. The 300 seconds of imaging yielded SNRs of about $35$ for NU and JH, and 30 for OE.  Using the calibration approach discussed in \S5.1, with Megaprime g' and r' filters and SNLS calibrated stars to calibrate the G and R values, we obtained calibration errors for (ZP$_1$,ZP$_2$) of (0.01,0.015), (0.01,0.14), (0.007,0.009) for NU, OE and JH respectively.  The positions of the SNe candidates in $CC-B$ space are shown in Fig. \ref{fig:AllArr}.   The photometric positions in $CC-B$ space are within one standard deviation from the corresponding synthetic positions, giving us confidence we are properly quantifying the SN flux through the SNACC filters with our photometry.

To determine the probability function a SN is a certain type and redshift, we employ the $\chi^2$ diagnostic of Eq. 8.  We make use of the sparse SN light curves to determine that our near-maximum SN templates are appropriate for the $\chi^2$ diagnostic.  The photometric redshifts from the host galaxy are $0.54\pm0.13$ for NU, $0.47\pm0.20$ for OE and $0.27\pm0.03$ for JH.  The probability distributions for each SN are shown in Fig. \ref{fig:FigNUDist}.  For SN NU, the probability that the SN is type Ia with $z=0.56$ is $\sim20\times$ the probability of any SNe II at any redshift.  There are negligible probability tails at $z=\pm0.2$ because the photometric redshift from the host galaxy is $\sigma_z=0.13$ and because of the NU's $CC-B$ position.  The measured redshift error uncertainty is thus $\sigma_z=0.01$ and the most probable redshift is within this error from the spectroscopic redshift.  For SN OE, the SN fell in the region of $CC-B$ space where there is a local degeneracy of $CC-B$ values.  The measurement strongly indicates that SN OE is type Ia, but with a poor photometric redshift $z=0.47\pm0.2$, it is difficult to determine the true redshift.  The spectroscopic redshift $0.42$ can be seen in one of the degenerate probability tails.  As seen in Fig. \ref{fig:AllArr}, the photometric and synthetic position are within one standard deviation of each other, so a more accurate measurement would not improve the classification.  While our 3-tooth prototype suffered from the degeneracy in the upper right quadrant of the $CC-B$ space, we were able to correct this in our 4-tooth solution, which will be used in the future and has no degenerate areas.  Lastly, for SN JH, the probability that the SN is a SN Ia at a given redshift is $<0.01$.  Although one can not say the redshift of the SN, the low probability for all SNe Ia would allow us to discriminate this SN from SNe Ia and remove it.

\section{Discussion}

While the use of multi-bandpass filters to measure a complex cross-correlation function is novel, such filters have had a prior use in astronomy.  They were used to monitor Blazars in multiple colors simultaneously (Wu et al. 2007).  The light passing through the different passbands may be differentially refracted by an objective prism and can then be spatially separated so there are multiple color images for each object obtained simultaneously. This system enables accurate tracing of contemporaneous color change of the blazars.  Surprisingly, because of the shape of certain Blazar spectra, the location of the bandpasses are very similar to those of the SNACC filters.

For our prototype 3-tooth design, we did not attempt to minimize the exposure time by maximizing the filter throughput.  We have begun work on this optimization and find that we can create 4-tooth filters, as shown in Fig.  \ref{fig:FigFilt4}, with effective widths of around $600~\AA$, which is more than half the width of broadband filters.  As seen in Fig. \ref{fig:FigCirc4}, this design avoids any local degeneracies, while still maintaining a purity of SNe Ia greater than $98\%$ and a redshift error $\sigma_z=0.01$, equivalent to what we can achieve with the prototype design.   

Because of gains in throughput, multiplexing and set-up, using SNACC filters is greater than an order of magnitude faster than long-slit spectroscopy.  For example, with SNACC filters designed for SuprimeCam at Subaru (Miyazaki et al. 2002), acquiring images of SNe (SNR $> 30$) is $\sim4$ times faster than acquiring spectra (with SNR $\sim30$, using the Subaru Exposure Time Calculator), and the expected space density of SNe Ia in a Pan-STARRS-like survey allows one to collect 4 to 8 SNe per pointing (with a field of view of 0.25 square degrees), so using SNACC filters is up to $\sim30$ times faster than spectroscopy.  

Spectra of high-z SNe ($z>0.5$) with a sufficient SNR for identification may be acquired on a large telescope (e.g., the Very Large Telescope (VLT), 8 m, see Appenzeller et al. 1998) in $<1$ hour, but the field of views for these telescopes' spectrographs is too small (e.g., FORS1 on VLT, 36 square arcmin) to allow any multiplex advantage from multi-slit spectroscopy.  On smaller telescopes (e.g., Magellan Baade and Clay Telescope 6.5 m, see Dressler 2004), it is more common to obtain spectra of SNe with redshifts $0.2<z<0.5$ (e.g., Foley 2009), and while the field of view may be larger (e.g., IMACS on Baade, 239 square arcmin), the viewing depth is not large enough to allow any significant multiplex advantage from multi-slit spectroscopy.

In future surveys, measurements with SNACC filters should yield a likelihood distribution for each individual SN candidate being type Ia within a measured redshift range.  Such a distribution is well incorporated into a Bayesian cosmology analysis.  While there are small tails in this probability distribution from the SNACC measurements of $\Delta z=\pm0.2$, an error in $z$ $=\pm0.2$ should be easily distinguishable in a Bayesian cosmological analysis of the Hubble diagram since the difference in distance for two SNe Ia with this redshift separation is $\sim 1.0$ mag.  Also, while there are certain transients, like SNe Ib-c, that are easily mistaken for SNe Ia, measurements with the SNACC filters will still provide the correct redshift of the SNe Ib-c.  With the right redshift, the rarity of these objects, and the known brightness difference between supernovae types, a Bayesian cosmological analysis should easily be able to cope with these contaminants (Kunz et al. 2006).  

Aside from speed, there are a number of advantages provided by the SNACC filters, which can be valuable depending on the application. The measurements with SNACC filters are much less sensitive to reddening and allow for a more precise determination of redshifts than usual photometric methods.  Image subtraction allows for a measurement free from the host contamination afflicting spectroscopy, which can be essential near the centers of hosts.  Indeed, a well known selection effect pertains at high redshift by the inability of long-slit spectroscopy to identify SNe Ia for cosmological analysis.  This bias may be lifted with SNACC filters.  And lastly, using the SNACC filters is independent of the color data used for subsequent distance determinations so there is no selection or interpretation bias (Kessler et al. 2009), and because it makes use of the spectral features its reliability is greater.  Although there are other approaches, like the use of multiple Dichroic stacks, that may be able to achieve the same goal of fast spectroscopy, SNACC filters are far more accessible and economical because they can be deployed in available filter slots.

\section{Summary}

To obtain the needed type and redshift information of candidate SN, we have developed a method of supernova observation which exploits the speed of photometric observations and much of the accuracy and precision of spectroscopic observations to determine the type of SN, and if Ia, its redshift.    Our approach is based on the use of multiple narrow passbands in a single filter.  Observations of supernovae through two of these filters effectively provide equivalent information as the evaluation of a spectral cross-correlation, typically used to measure the redshift and type of a SN from a low SNR spectrum.  Simulating the use of SuperNovae Cross Correlation Filters (SNACC) filters as a follow-up tool, we can obtain a sample of SNe Ia which is $\sim98\%$ pure with redshifts of individual precision $\sigma_z=0.01$ at a rate up to $\sim30$ times faster than typical high-z SN spectroscopy, a speed advantage which could greatly benefit future wide-area transient surveys. 

Prototype SNACC filters were built and tested on Magellan with LDSS-3 and used to follow-up three SNLS candidates. We acquired images with the SNACC filters as well as spectra of each candidate and we were successful in classifying the three candidates.   In the future we plan to apply this new technology for the next generation of supernova surveys and try to improve the precision of the supernova sample and its constraining power for dark energy.

\acknowledgments
We thank Alan Uomoto at Carnegie Observatories for all of his assistance, and the JHU Optics Lab for their help in measuring the filter transmission functions.  The procurement of the filters was made possible through the support of the STScI Director's Discretionary Research Fund.  We wish to thank the Supernova Legacy Survey (SNLS) collaboration, and in particular Kathy Perrett and Alex Conley, for providing SNe candidates for follow-up. This paper is based on data obtained at the 6.5 meter Magellan Telescopes located at Las Campanas Observatory, Chile.  This research has made use of the SDSS DR7 database.  

Funding for the SDSS and SDSS-II has been provided by the Alfred P. Sloan Foundation, the Participating Institutions, the National Science Foundation, the U.S. Department of Energy, the National Aeronautics and Space Administration, the Japanese Monbukagakusho, 
the Max Planck Society, and the Higher Education Funding Council for England. The SDSS Web site is http://www.sdss.org/. 
The SDSS is managed by the Astrophysical Research Consortium for the Participating Institutions. The Participating Institutions are the American Museum of Natural History, Astrophysical Institute Potsdam, University of Basel, Cambridge University, Case Western 
Reserve University, University of Chicago, Drexel University, Fermilab, the Institute for Advanced Study, the Japan Participation Group, Johns Hopkins University, the Joint Institute for Nuclear Astrophysics, the Kavli Institute for Particle Astrophysics and Cosmology, the Korean Scientist Group, the Chinese Academy of Sciences (LAMOST), Los Alamos National Laboratory, the Max-Planck-Institute for Astronomy (MPIA), the Max-Planck-Institute for Astrophysics (MPA), New Mexico State University, Ohio State University, University of Pittsburgh, University of Portsmouth, Princeton University, the United States Naval Observatory, and the University of Washington.

Facilities: Magellan: Clay (LDSS-3), CFHT.

\clearpage



\begin{deluxetable}{l l c  c }
\tabletypesize{\scriptsize}
\tablecaption{Usage Scenarios}
\tablewidth{0pt}
\tablehead{\colhead{Scenario} & \colhead{Details} & \colhead{Results} & \colhead{} \\ 
\colhead{} & \colhead{} & \colhead{SNe Ia Correctly Classifed (\%)} & \colhead{$\sigma_z$} } 
\startdata
 Host Galaxy, Light Curve  & $\sigma_z$=0.10, $\sigma_z/(1+z)\approx 0.06$\tablenotemark{a}, & 80.0 (5.00) &  0.090 (0.010)  \\  
  SNACC, Host Galaxy w/ Photo-z, no LC   & $\sigma_z$=0.10 & 94.5(0.2)  & 0.015 (0.003)  \\ 
  
  SNACC, Host Galaxy w/ Spec. z, no LC   & $\sigma_z$=0.01  & 95.6 (0.3)  & ... \\   
  
  SNACC, No Host Galaxy, w/ LC &  $\sigma_z/(1+z)\approx 0.06$\tablenotemark{a}, &  97.0 (0.2) & 0.014 (0.002)  \\ 
  & known age, stretch, reddening & & \\

  SNACC, Host Galaxy w/ Photo-z, w/ LC   & $\sigma_z$=0.10, $\sigma_z/(1+z)\approx 0.06$\tablenotemark{a}, & 98.2 (0.1) &  0.012 (0.002)  \\ 
  & known age, stretch, reddening && \\

  SNACC, Host Galaxy w/ Spec-z, w/ LC   & $\sigma_z$=0.01, & 99.0 (0.1) &  ...  \\ 
& known age, stretch, reddening & & \\

\enddata
\tablenotetext{a}{These photometric redshifts, derived from the broadband colors of the SN, are dependent on whether the SN is known to be a SN Ia.}
\end{deluxetable}



\begin{deluxetable}{lcccccl}
\tabletypesize{\footnotesize}
\tablecaption{SN Observation Log. \label{tab_log}}
\tablewidth{0pt}
\tablehead{
\colhead{Source} &
\colhead{Coordinates [J2000.0] of SN} &
\colhead{Imag./Spec.} &
\colhead{UT Date} &
\colhead{$t_{int}$ (s)} &
\colhead{Offset [arcsec]$^{a}$} &
\colhead{Airmass}  
}
\startdata
SN~08D2nu&$09^h58^m32^s.778\quad +02^{\circ}36'05''.98$ &Imag.&2008-05-03&300&0.54E\quad2.03N&1.214\\
&&Spec. &2008-05-04&$2\times1200$&&1.220\\

SN~08D2oe&$10^h01^m14^s.136\quad +02^{\circ}04'16''.99$ &Imag.&2008-05-03&300&...{$^b$}&1.434\\
& &Spec.&2008-05-09&$2\times1800$&~&1.170\\

SN~08D2jh&$09^h59^m31^s.661\quad +02^{\circ}14'51''.93$ &Imag.&2008-05-05&300&0.01W\quad0.18N&$1.170$\\
&&Spec.&2008-05-04&$2\times900$&&$1.460$\\

\enddata
\tablenotetext{a}{SN positional offset from nucleus of the host galaxy}
\tablenotetext{b}{SNR$<5$ of host galaxy for all data}

 \end{deluxetable}
 

\begin{deluxetable}{lcccccl}
\tabletypesize{\footnotesize}
\tablecaption{Spectrophotometric Standards Imaging Observation Log. \label{tab_log}}
\tablewidth{0pt}
\tablehead{
\colhead{Source$^{a}$} &
\colhead{Spectral Type} &
\colhead{UT Date} &
\colhead{$t_{int}$} &
\colhead{Airmass} & 
\colhead{g' } &
\colhead{r'}  \\
\colhead{} & \colhead{} & \colhead{}  & \colhead{(s)} & \colhead{} &\colhead{(mag)} & \colhead{(mag)}
}
\startdata

SDSS J141101.76+010418.4&G8V Star&2008-05-03&180&1.490&$17.19 \pm 0.01$&$16.67\pm0.01$\\
SDSS J125857.35+145019.7&M4V Star&2008-05-03&180&1.157&$18.97 \pm 0.01$&$17.40\pm0.01$\\

\enddata
\tablenotetext{a}{The coordinates are given inside the source name, in HHMMSS of J2000.} 

 \end{deluxetable}



\begin{deluxetable}{ccccccccc}
\tabletypesize{\scriptsize}
\tablecaption{Main SNe Characteristics \label{SNsample} }
\tablewidth{0pt}
\tablehead{\colhead{SN} &\colhead{Type}& \colhead{Age$^{a}$}&\colhead{g'} &\colhead{r'} & \colhead{Host Gal. Photo-z$^{b}$} & \colhead{SNLS photo-z } &\colhead{ Spec. z$^{d}$}\\ 
\colhead{} & \colhead{} & \colhead{(days)} & \colhead{(mag)}  & \colhead{(mag)} & \colhead{} & \colhead{(if SN Ia)$^{c}$} & \colhead{} } 
\startdata
NU&Ia&+1&23.4&22.5&$0.54 \pm 0.13$&0.42&0.56\\
OE&Ia&-3&23.3&22.9&$0.47 \pm 0.20$&0.42&0.42\\
JH&IIP&+30&23.3&22.5&$0.27 \pm 0.03$&0.36&0.22\\

\enddata
\tablenotetext{a}{Relative to the epoch of B-band maximum}
\tablenotetext{b}{Redshift of the host galaxy, as determined by the Sloan Digital Sky Survey (SDSS) pipeline for SDSS galaxies}
\tablenotetext{c}{SNLS calculated redshift using broadband colors of SN}
\tablenotetext{d}{SNID redshift from our LDSS-3 spectra}
\end{deluxetable}


   
\pagebreak

\begin{figure}[h]
\begin{center}
\includegraphics[viewport=.5in 1.5in 9.0in 10.0in, scale=0.75]{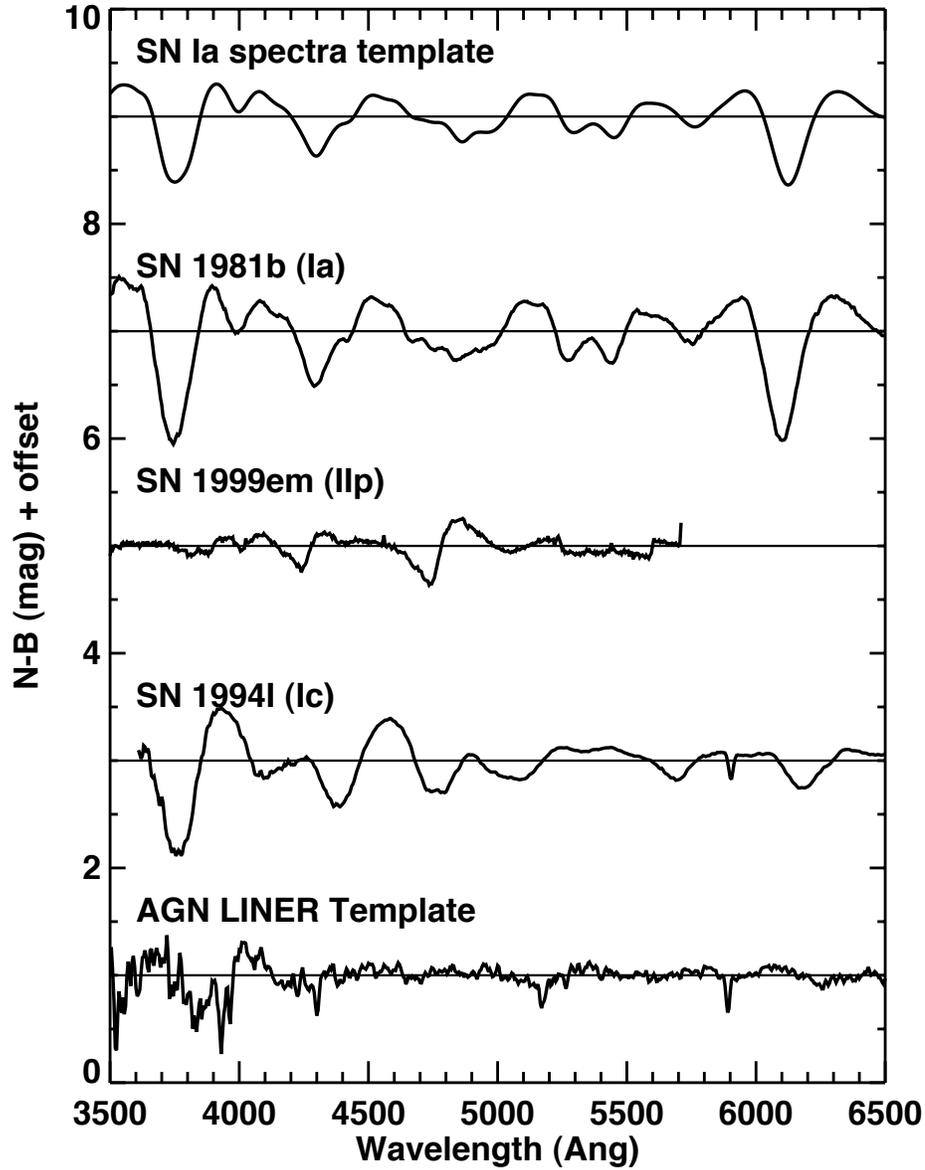}
\end{center}
\caption{
Rest-frame wavelength spectra of supernova types, as well as an AGN template: SN Ia template (Hsiao et al. 2008), SN 1981b (Branch et al. 1983), SN 1999em (Baron et al. 2000), SN 1994I (Clocchiatti et al. 1996) and AGN LINER template (Francis et al. 1991).    The spectra are calibrated on the y-axis by $2.5 \times \log (F_{N}/F_B)$, where $F_{N}$ is the flux in a one-angstrom narrow-band filter and $F_B$ is the flux in a $500~\AA$ filter centered around the narrow filter.  Offsets, shown by the solid lines, are added to the values.  Compared to spectra like that of the SN IIP, or the AGN, the SNe Ia has very deep and wide features. }
\label{fig:FigSpec}
\end{figure}
\pagebreak


\begin{figure}[h]
\begin{center}
\includegraphics[viewport=1.0in .5in 9.0in 9.0in, scale=0.75]{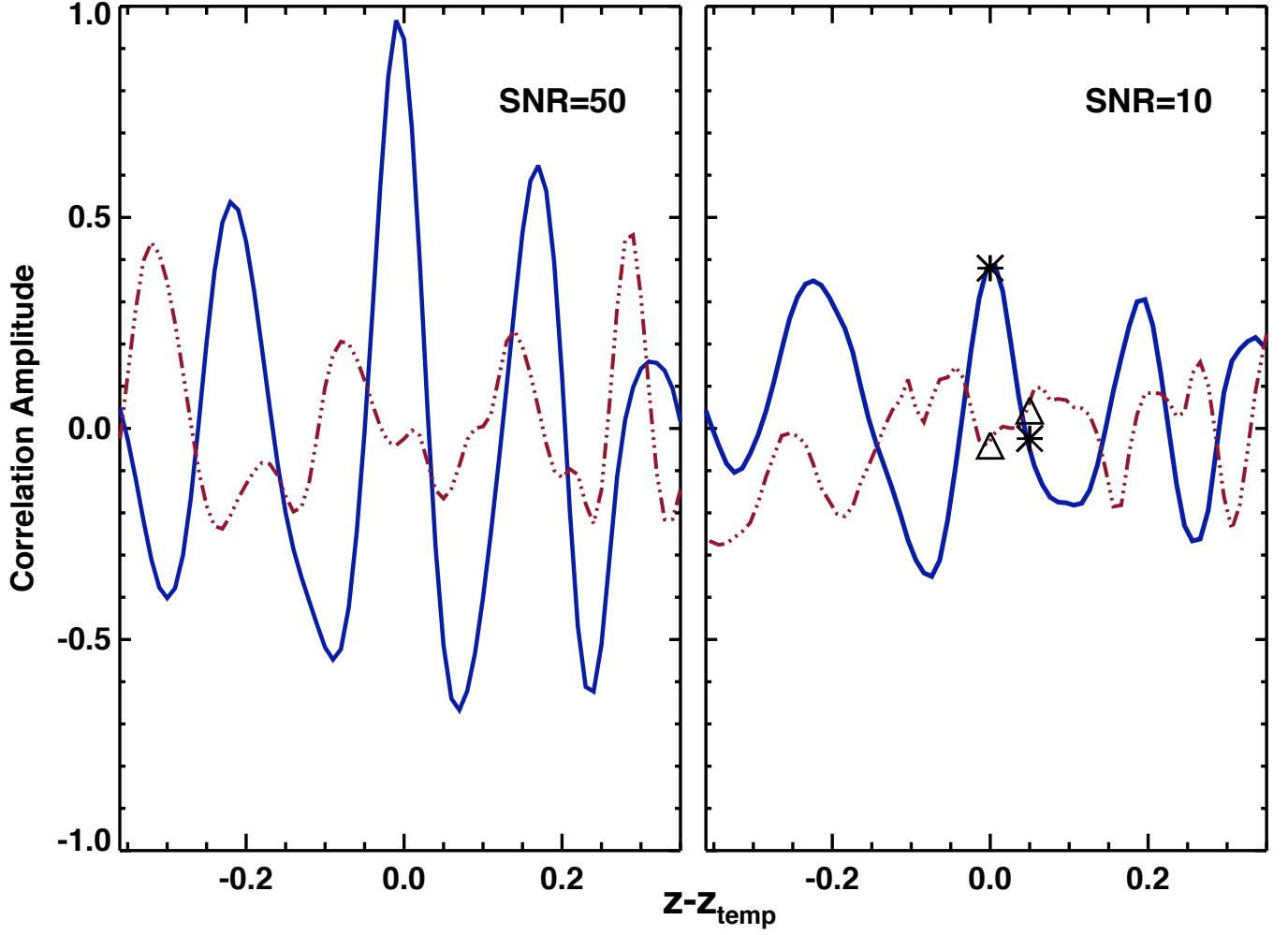}
\end{center}
\caption{
(left) The correlation function of a \textit{high} SNR spectrum with a SN Ia template (solid) and SN II template (dotted).  The peak value of the correlation curve allows one to determine the true redshift, in this case $z-z_{temp}=0$.  (right)  The correlation function of a \textit{low} SNR spectrum with a SN Ia template (solid) and SN II template (dotted).  Here, a discrete set of redshifts, $z-z_{temp}=0$, and $z-z_{temp} = \pm0.2$ is preferred. The marks in the right panel demonstrate how with two measurements of the correlation curve, one can discriminate the type of SN, and for SNe Ia, determine its redshift to $\sigma_z\approx0.01$. }
\label{fig:FigFourB}
\end{figure}

\begin{figure}[h]
\begin{center}
\includegraphics[viewport=1.0in 1.0in 11.0in 8.0in, scale=0.75]{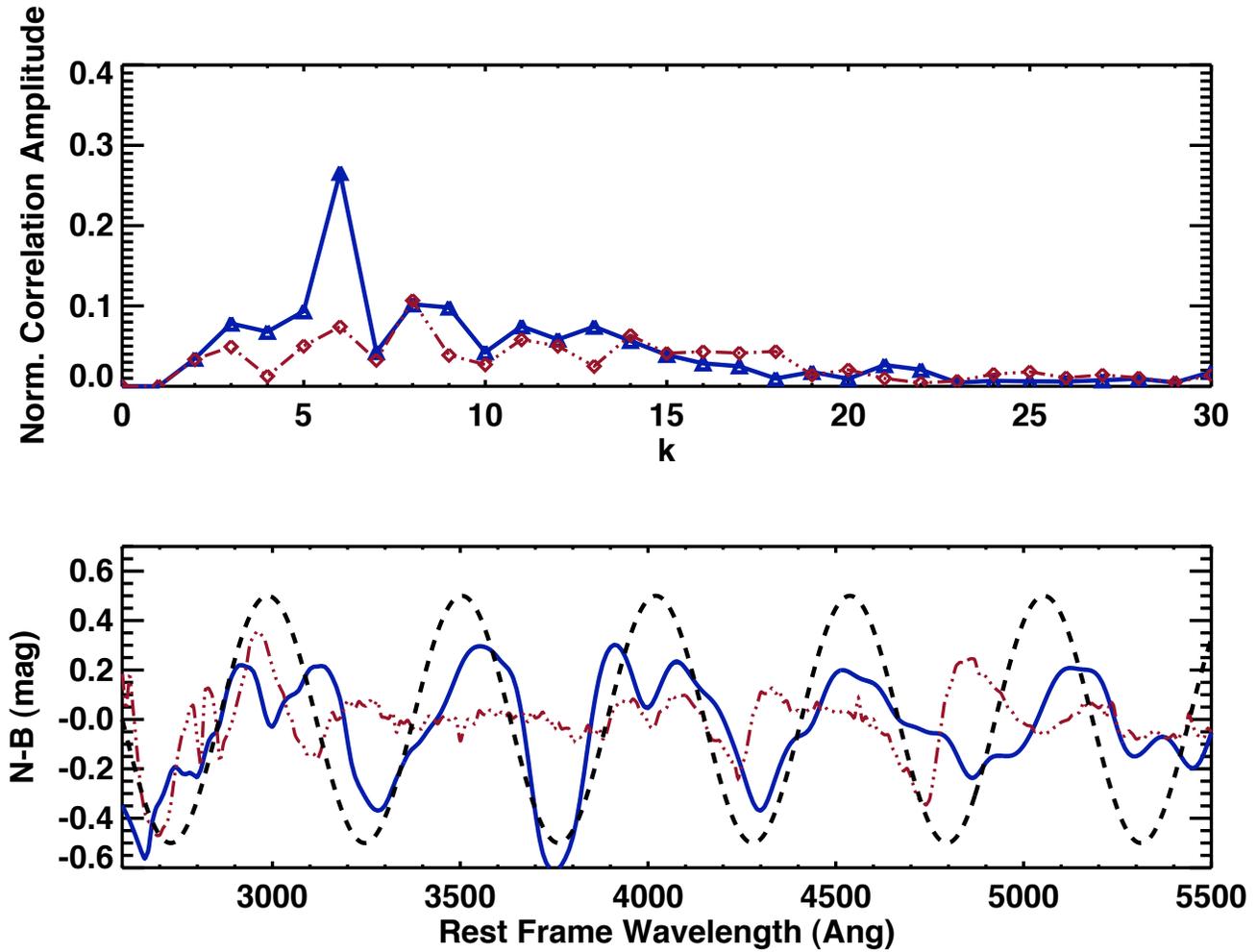}
\end{center}
\caption{
The fourier transform of a SN Ia (z=0, peak brightness) spectral template (solid-triangles) and SN II (z=0, peak brightness) spectral template (dashed-squares) in the top panel.  The SN Ia spectrum shows the majority of its correlation amplitude at $k=6$, or $\sim 500~\AA$, while the correlation amplitude for the SN II spectrum at $k=6$ is much weaker.  In the lower panel, a sine curve (dashed line) with wavelength $500~\AA$ is overplotted on a SN Ia (solid) and SN II spectra (dotted).}
\label{fig:FigFourIa}
\end{figure}

\begin{figure}[h]
\begin{center}
\includegraphics[viewport=1.0in 1.0in 11.0in 8.0in, scale=0.75]{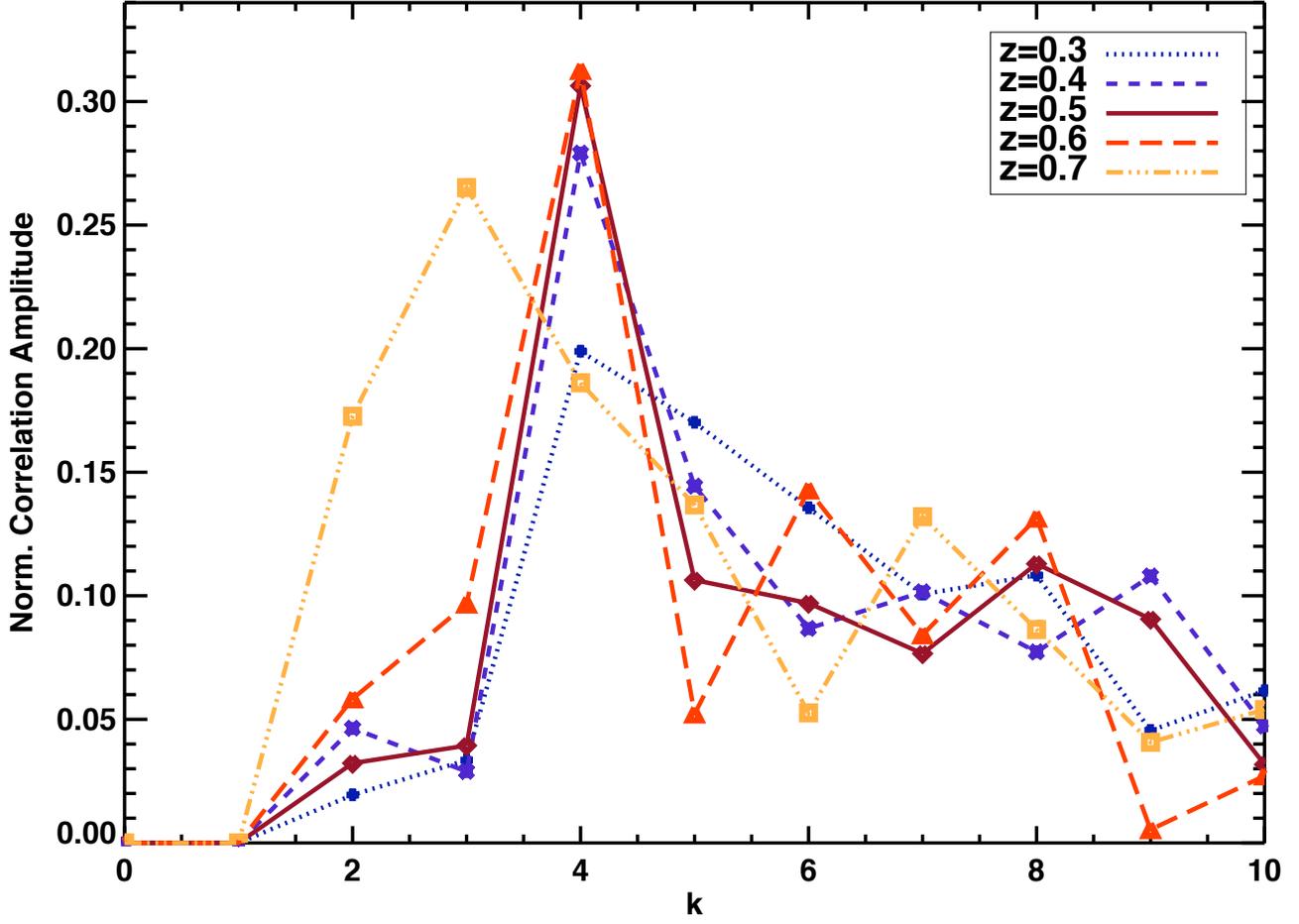}
\end{center}
\caption{
Fourier transforms of SNe Ia spectra in the observed frame at redshifts $0.3 < z < 0.7$.  For each redshifted spectrum, there is a high correlation amplitude at $k=4$, which is equivalent to a wavelength separation of $775~\AA$.  The peak wavenumber depends on the redshift $z_s$ such that $k_p=6/(1+z_s)$, but changes little over a modest fractional change in $(1+z_s)$.}
\label{fig:FigFourAll}
\end{figure}

\pagebreak
\begin{figure}[h]
\begin{center}
\includegraphics[viewport=1.5in 1.0in 11.0in 8.0in, scale=0.75]{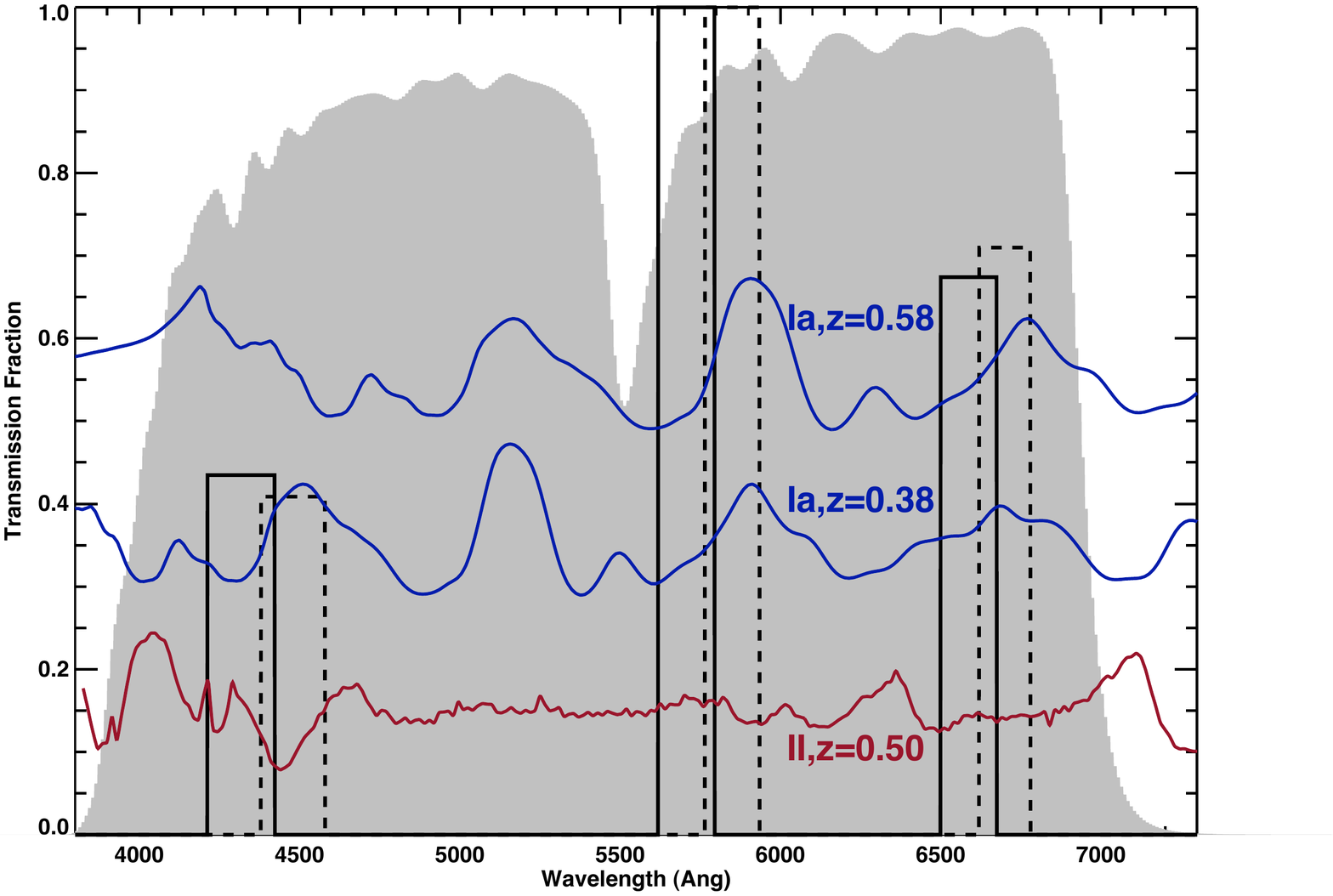}
\end{center}
\caption{
The optimized three-tooth filter design, with one filter portrayed as a solid line, and the other as a dashed line.  The g' and r' Sloan filters are seen in gray. SN Ia spectra at $z=0.38,~0.58$ are overplotted to illustrate how the SNACC filter passbands correlate with the SNe Ia spectral features.  A SN IIP spectrum at $z=0.50$ is also shown to demonstrate little correlation between the SN II features and the filter teeth.}
 \label{fig:FigFilt3}
\end{figure}
\pagebreak
\begin{figure}[h]
\begin{center}
\includegraphics[viewport=1.0in .5in 11.0in 10.0in, scale=0.75]{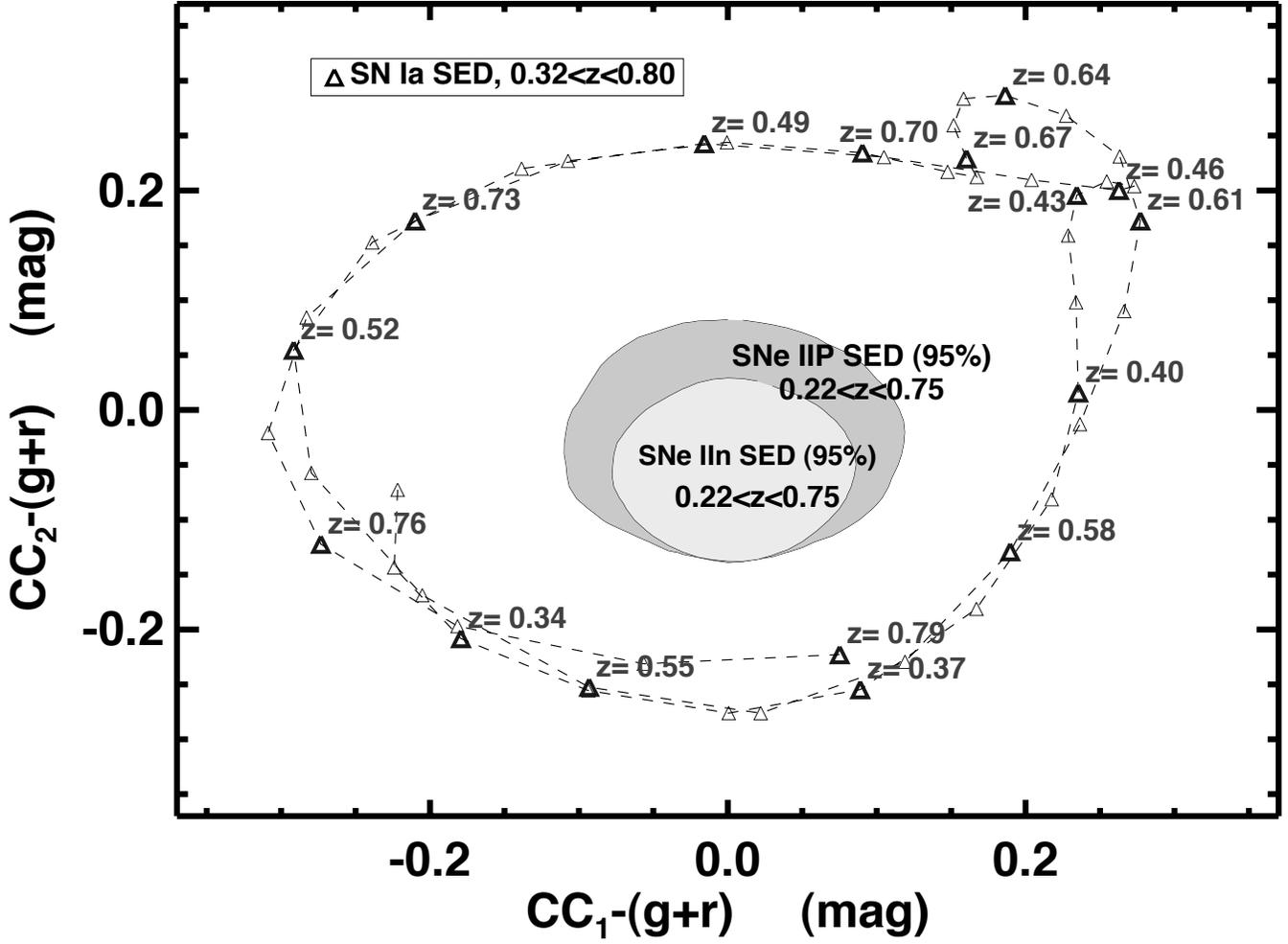}
\end{center}
\caption{
The $CC-(g+r)$ diagram for the prototype design of the SNe Ia (outside) and the SNe II (inside).  The SNe II redshifts cluster in the center of the diagram, and the SN Ia redshifts have an average radius $\approx0.25$ mag.  For measurement errors $\sigma_{1,2}=0.04$, the SNe Ia are separated from the SNe II in $CC-(g+r)$ space by $3-5 \times \sigma_{1,2}$, allowing for a SNe Ia sample purity of $\sim98\%$.  }
 \label{fig:FigCirc3}
\end{figure}
\pagebreak

\pagebreak
\begin{figure}[h]
\begin{center}
\includegraphics[viewport=1.0in .5in 11.0in 8.0in, scale=0.75]{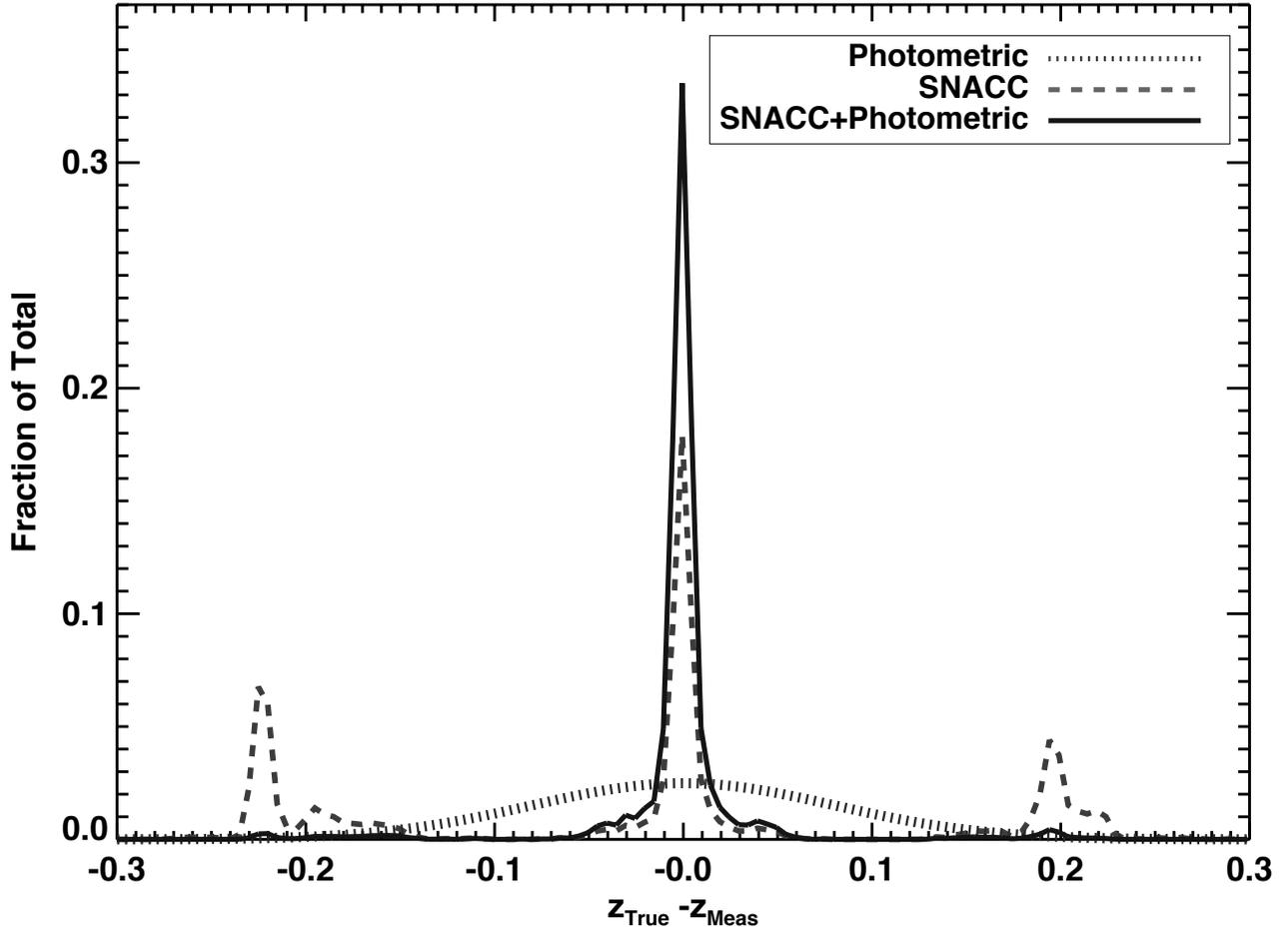}
\end{center}
\caption{
Histogram of redshift estimation with only a photometric redshift (dotted), only SNACC filters (solid) and the combination of SNACC filters with a photometric redshift (dashed).  The tails in the population (3\% for SNACC+Photometric, 45\% for just SNACC) of redshift errors at $\pm 0.2$ are due to the degeneracy of the redshift in the correlation curves.  Overall, the combination of SNACC filters with the photometric redshift yields a main redshift error of $\sigma_z \approx 0.01$.}
\label{fig:FigHist}
\end{figure}


\clearpage

\pagebreak
\begin{figure}[h]
\begin{center}
\includegraphics[viewport=0.5in 2.5in 11.0in 10.0in, scale=0.75]{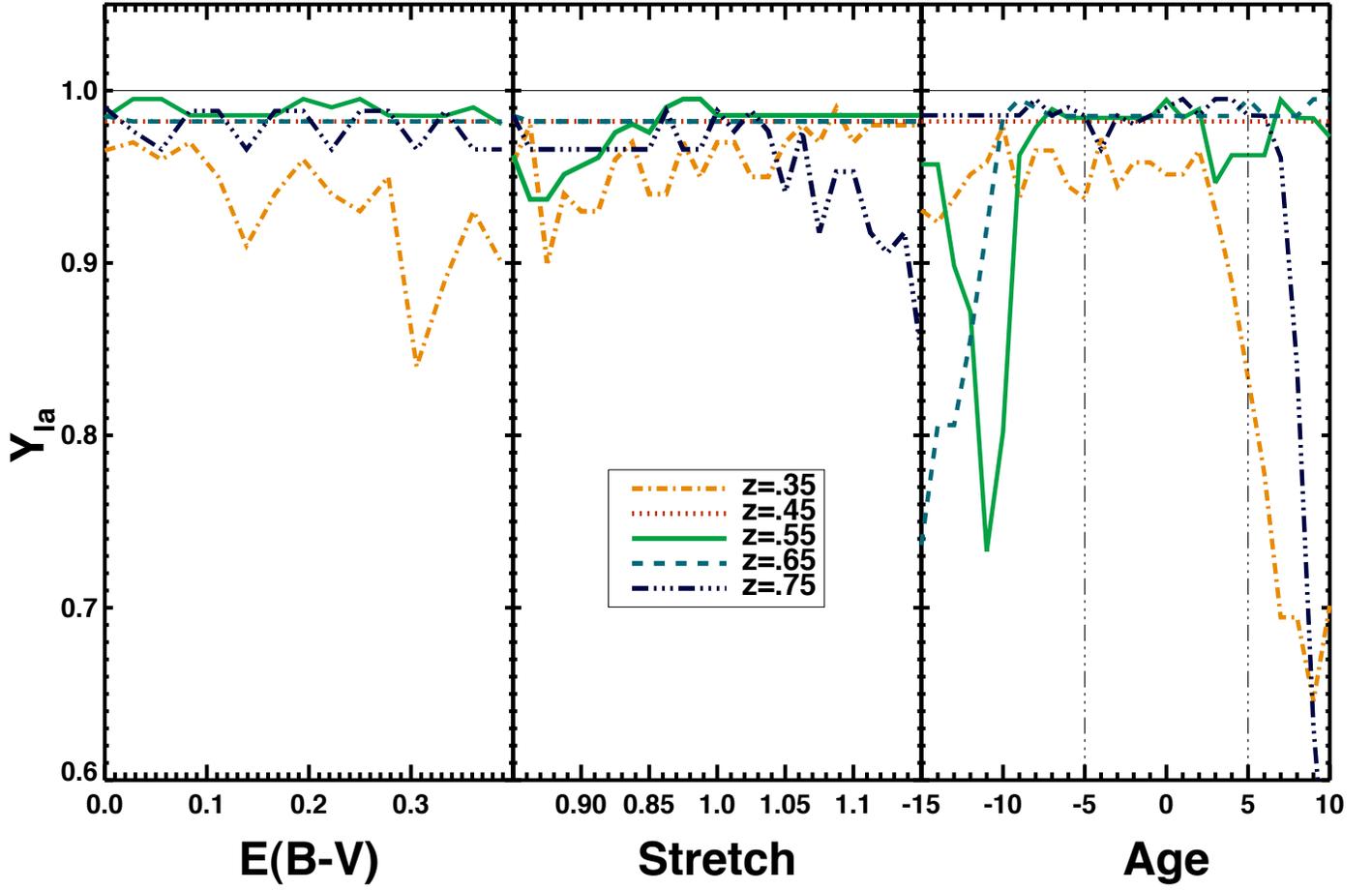}
\end{center}
\caption{
Effects of reddening, stretch and age on the resulting fraction, $Y_{Ia}$, of SNe Ia correctly classified (type) for various redshifts.  Reddening has very little effect on classification, except for the lowest redshifts, while stretch has a similarly small effect except for the highest redshifts.  The filters were optimized for use around peak ($\pm 5$ days), and the classification is generally $>90\%$ this range. }
\label{fig:FigFactors}
\end{figure}

\pagebreak
\begin{figure}[h]
\begin{center}
\includegraphics[viewport=.5in 2.5in 11.0in 10.0in, scale=0.75]{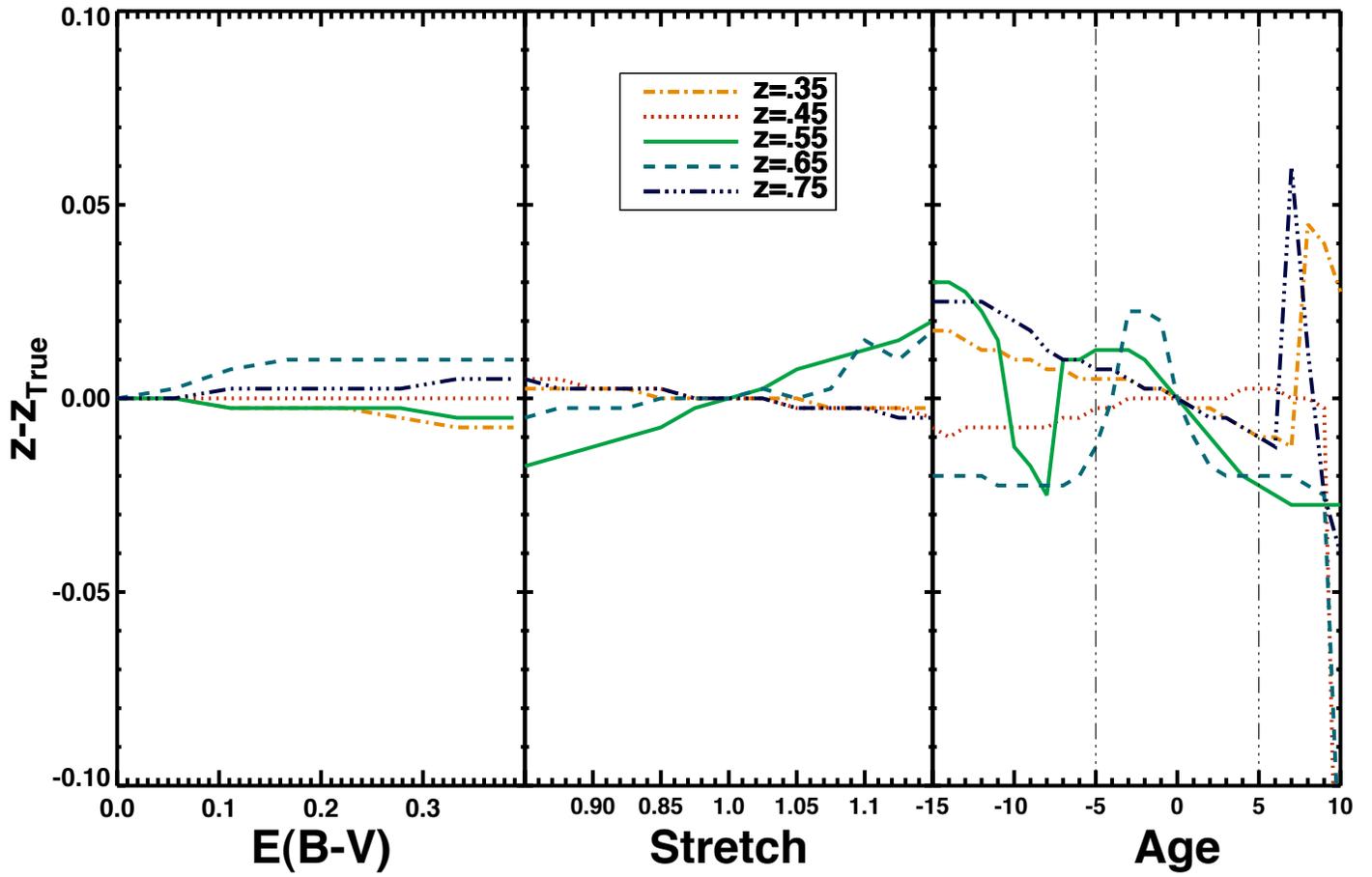}
\end{center}
\caption{
Effects of ignorance of reddening, stretch and age on the errors in determined redshift.  Variations in reddening and stretch have a very small effect, producing redshift errors $<0.02$.  The filters were optimized for use around peak ($\pm 5$ days), and the redshift error for this age range is mostly $<0.02$.  Knowledge of the reddening, stretch and age would eliminate these errors.}
\label{fig:FigFactorsR}
\end{figure}

\pagebreak
\begin{figure}[h]
\begin{center}
\includegraphics[viewport=2.5in .5in 8.0in 10.0in, scale=0.75]{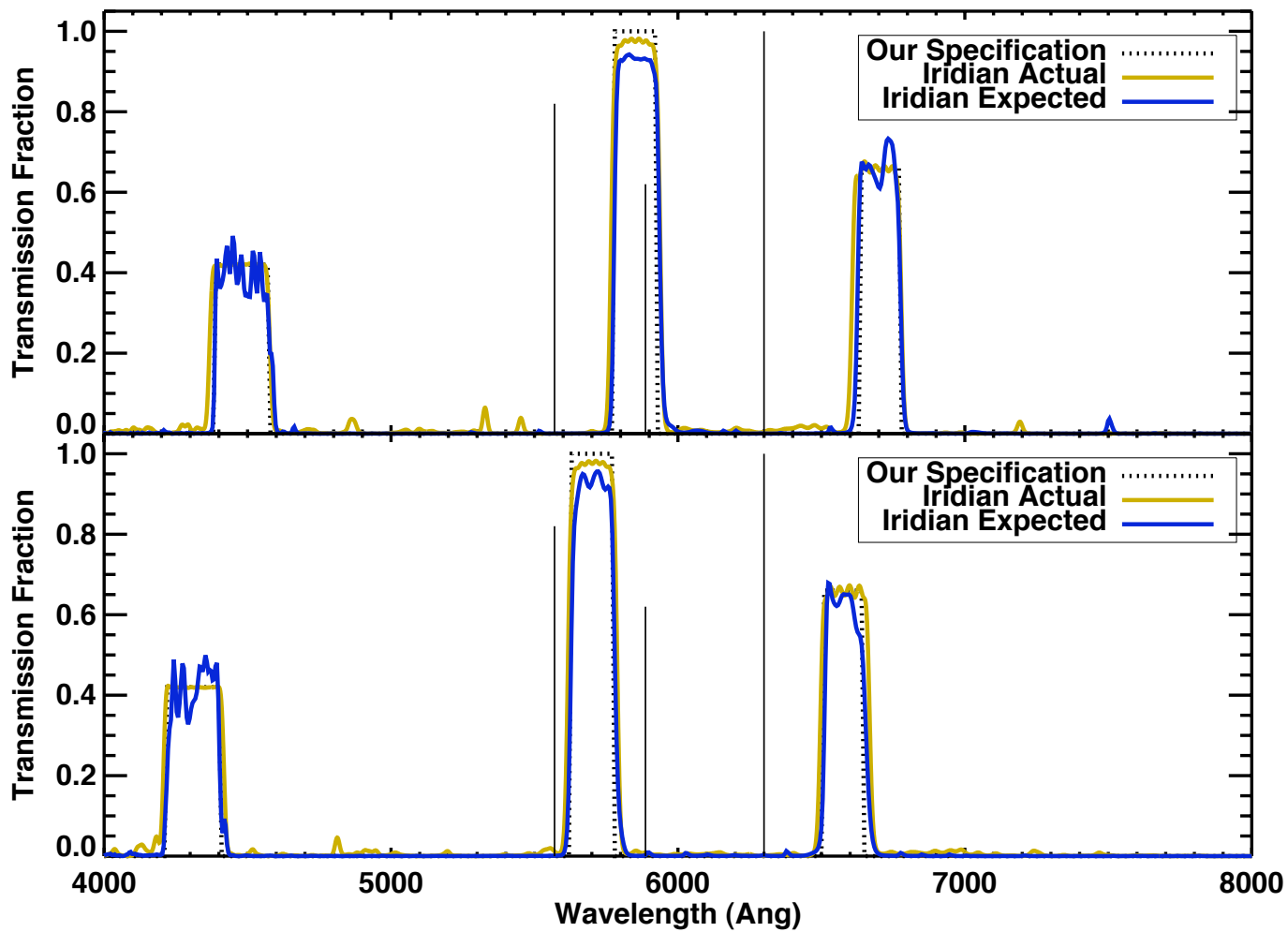}
\end{center}
\caption{
The bandpasses, Iridian's expected design, and Iridian's actual design, as well as the strongest skylines ([OI], Na D with relative strengths as found at Magellan) in the optical range.  Iridian was able to build filters with bandpasses of an accuracy better than $5\%$ of the specified widths, heights and positions of the teeth.}
\label{fig:FigIridian}
\end{figure}

\clearpage
\pagebreak
\begin{figure}[h]
\begin{center}
\includegraphics[viewport=1.0in 2.0in 10.0in 10.0in, scale=0.75]{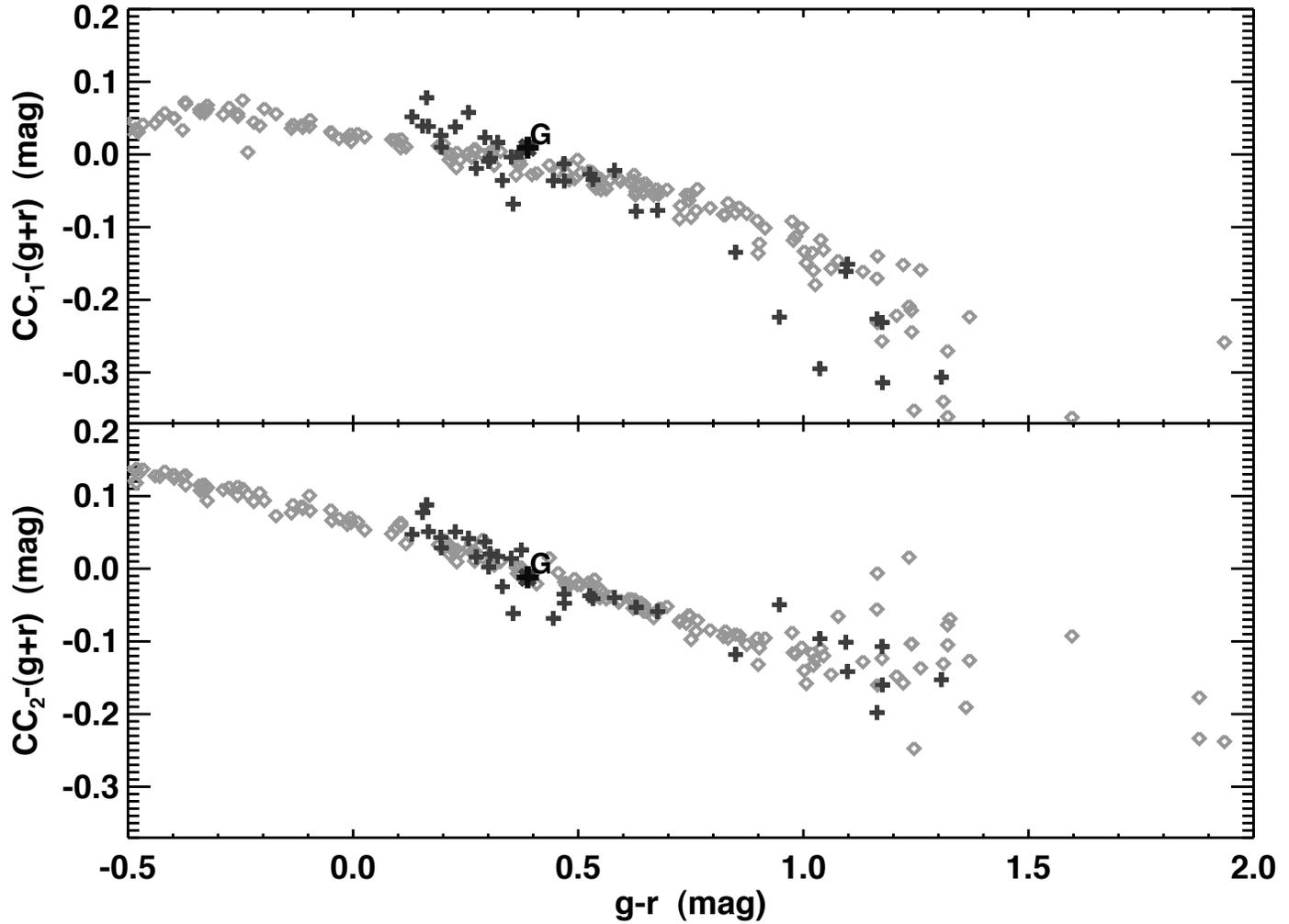}
\end{center}
\caption{
The calibration of the stars (crosses) in the G8V field with the set of Gunn-Stryker stars (diamonds), after the $CC-(g+r)$ offset is incorporated.  The correct $CC-(g+r)$ offset is found by fitting all of stars in the field with $g-r \le 1.0$ to the line formed of the Gunn-Stryker stars, and the calibration error is due to the differences between the set of stars in the field and the Gunn-Stryker set.  For this set of stars, the $CC_1-(g+r)$ offset is $0.33\pm0.01$ mag and the $CC_2-(g+r)$ offset is $0.24\pm0.005$ mag, where the errors here are folded into the entire photometric error of the classification process.  Once the stars have been calibrated, the $CC-(g+r)$ value of the G8V star can be placed in the $CC-(g+r)$ diagnostic space.}
\label{fig:FigGShift}
\end{figure}

\pagebreak
\begin{figure}[h]
\begin{center}
\includegraphics[viewport=1.0in 2.0in 10.0in 10.0in, scale=0.75]{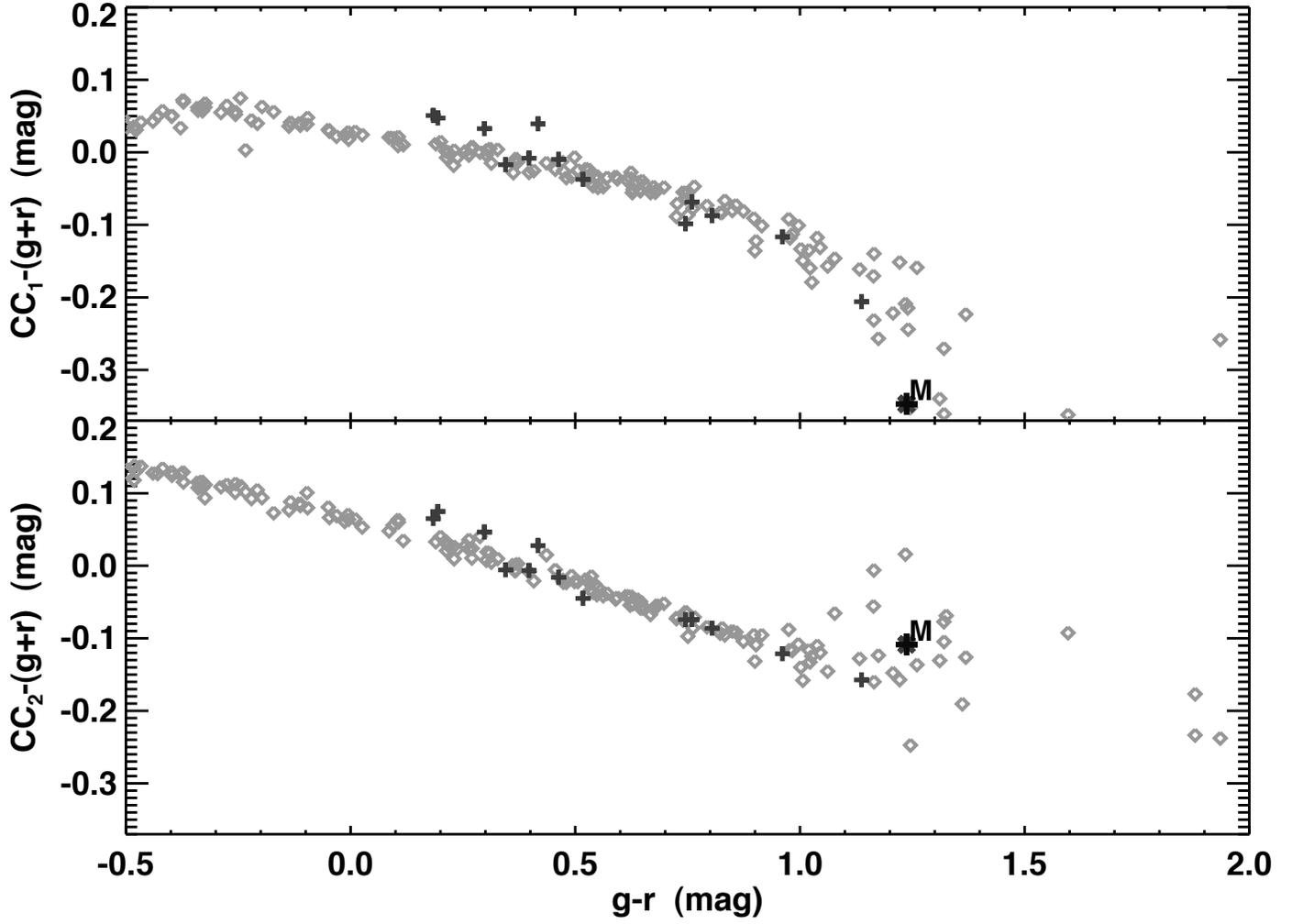}
\end{center}
\caption{
Similar to Fig. \ref{fig:FigMShift}, the calibration of the stars (crosses) in the M4V field with the set of Gunn-Stryker stars (diamonds), after the $CC-(g+r)$ offset is incorporated.   For this set of stars, the $CC_1-(g+r)$ offset is $0.33\pm0.015$ mag and the $CC_2-(g+r)$ offset is $0.22\pm0.01$ mag.  }
\label{fig:FigMShift}
\end{figure}

\pagebreak
\begin{figure}[h]
\begin{center}
\includegraphics[viewport=1.0in 2.5in 10.0in 10.0in, scale=0.75]{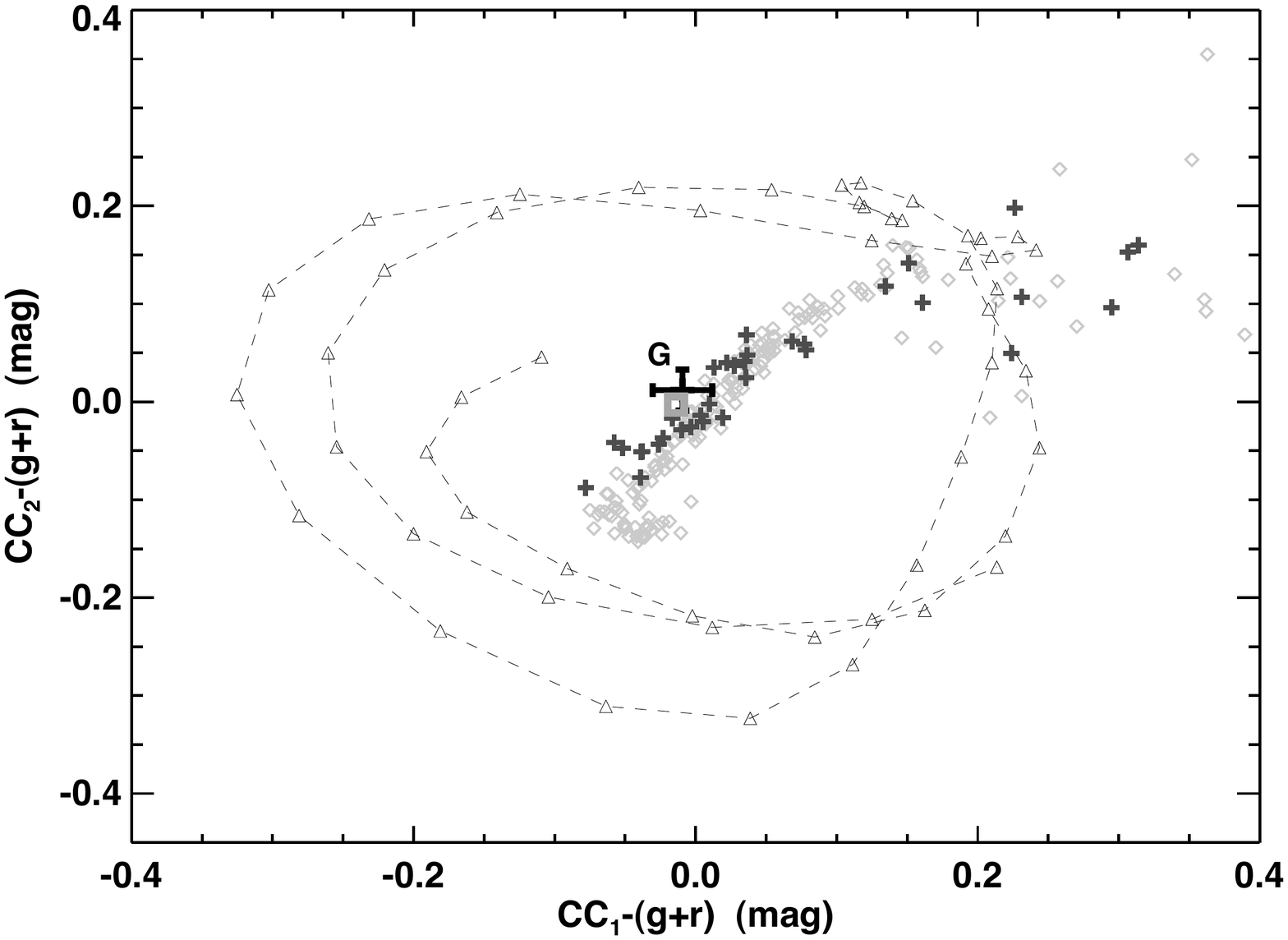}
\end{center}
\caption{
Combining Fig. \ref{fig:FigGShift}, the magnitudes of the stars (crosses) in the G8V field in $CC-(g+r)$ space, along with the Gunn-Styker stars (diamonds) and SNe Ia (triangles).  The photometric value (with error bars) of the G8V star is within one standard deviation of its synthetic position (highlighted square).}
\label{fig:FigMArr}
\end{figure}

\pagebreak
\begin{figure}[h]
\begin{center}
\includegraphics[viewport=1.0in 2.5in 10.0in 10.0in, scale=0.75]{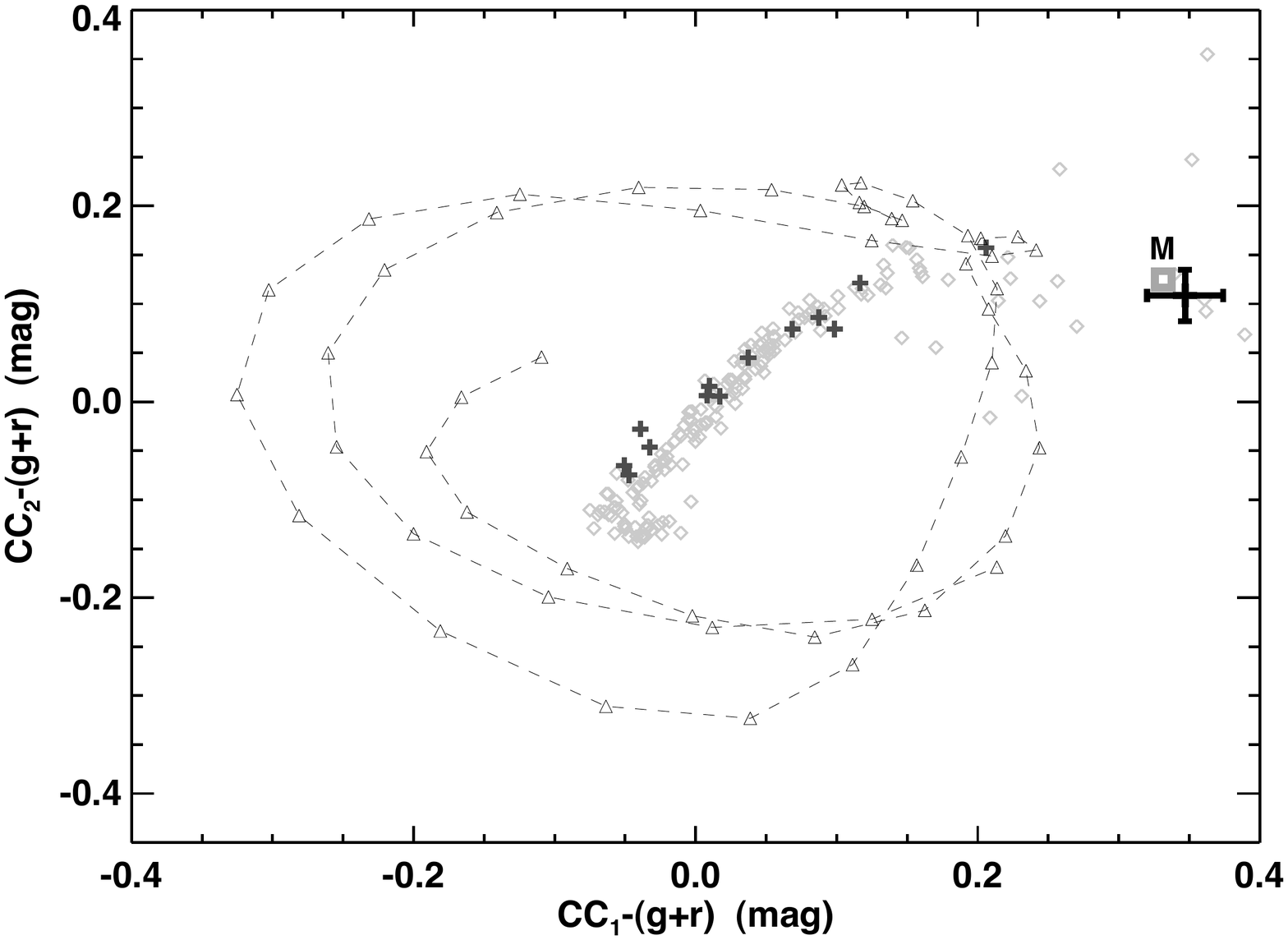}
\end{center}
\caption{
Combining Fig. \ref{fig:FigMShift}, the magnitudes of the stars (crosses) in the M4V field in $CC-(g+r)$ space, along with the Gunn-Styker stars (diamonds) and SNe Ia (triangles).  The photometric value (with error bars) of the M4V star is within one standard deviation of its synthetic position (highlighted square).}
\label{fig:FigGArr}
\end{figure}

\pagebreak
\begin{figure}[h]
\begin{center}
\includegraphics[viewport=1.0in 2.5in 10.0in 10.0in, scale=0.75]{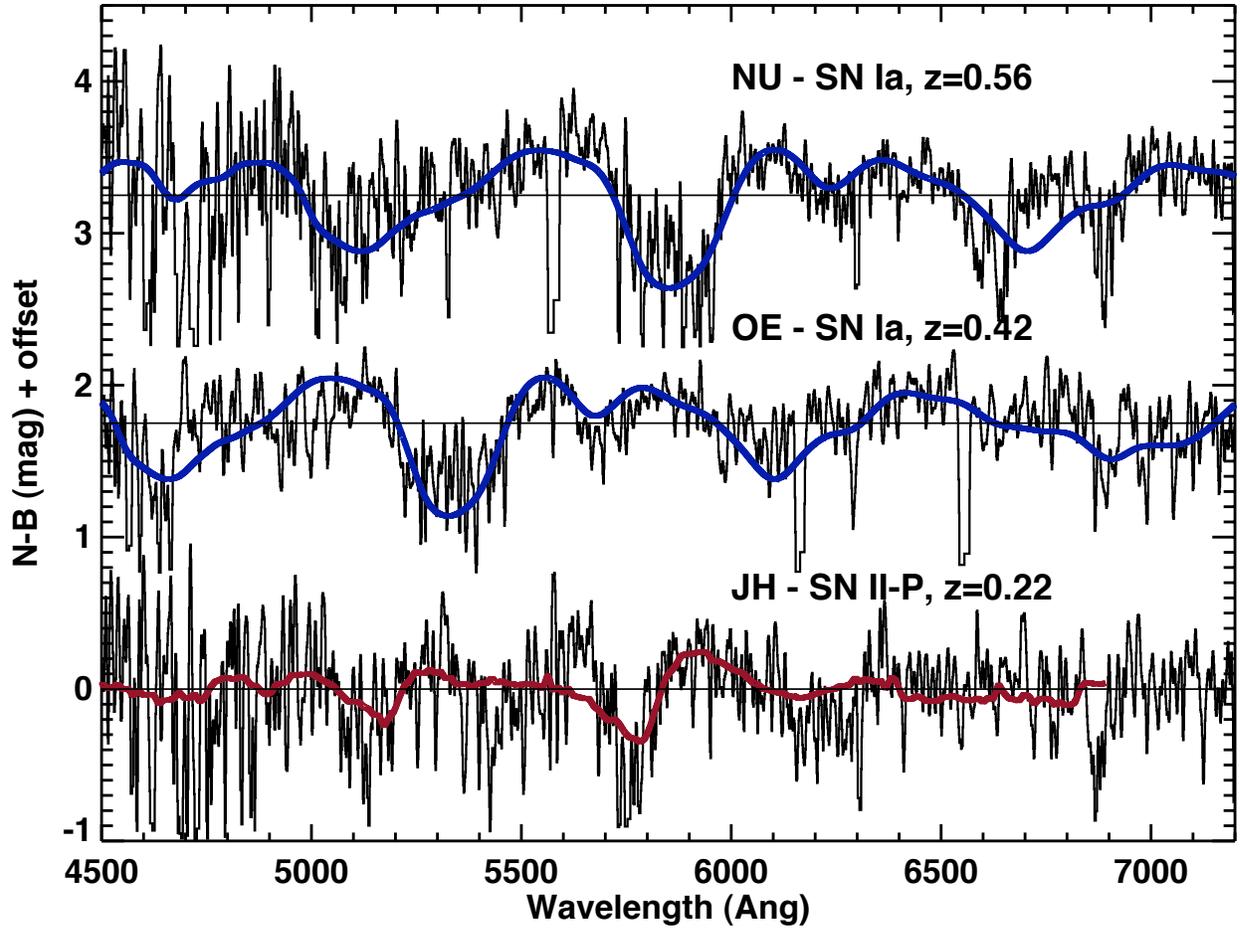}
\end{center}
\caption{
Observed Spectra for NU, OE and JH with SN templates over-plotted.  Using the SNID program, the supernovae are classified as: NU, a SN Ia with $z=0.56\pm0.01$; OE, a SN Ia with $z=0.42\pm0.01$; JH, a SN IIP with $z=0.22\pm0.01$.  These correlations have SNID rlap values of  0.63, 0.68 and 0.74 respectively.}
\label{fig:FigAllSpec}
\end{figure}

\pagebreak
\begin{figure}[h]
\begin{center}
\includegraphics[viewport=1.0in 2.5in 10.0in 10.0in, scale=0.75]{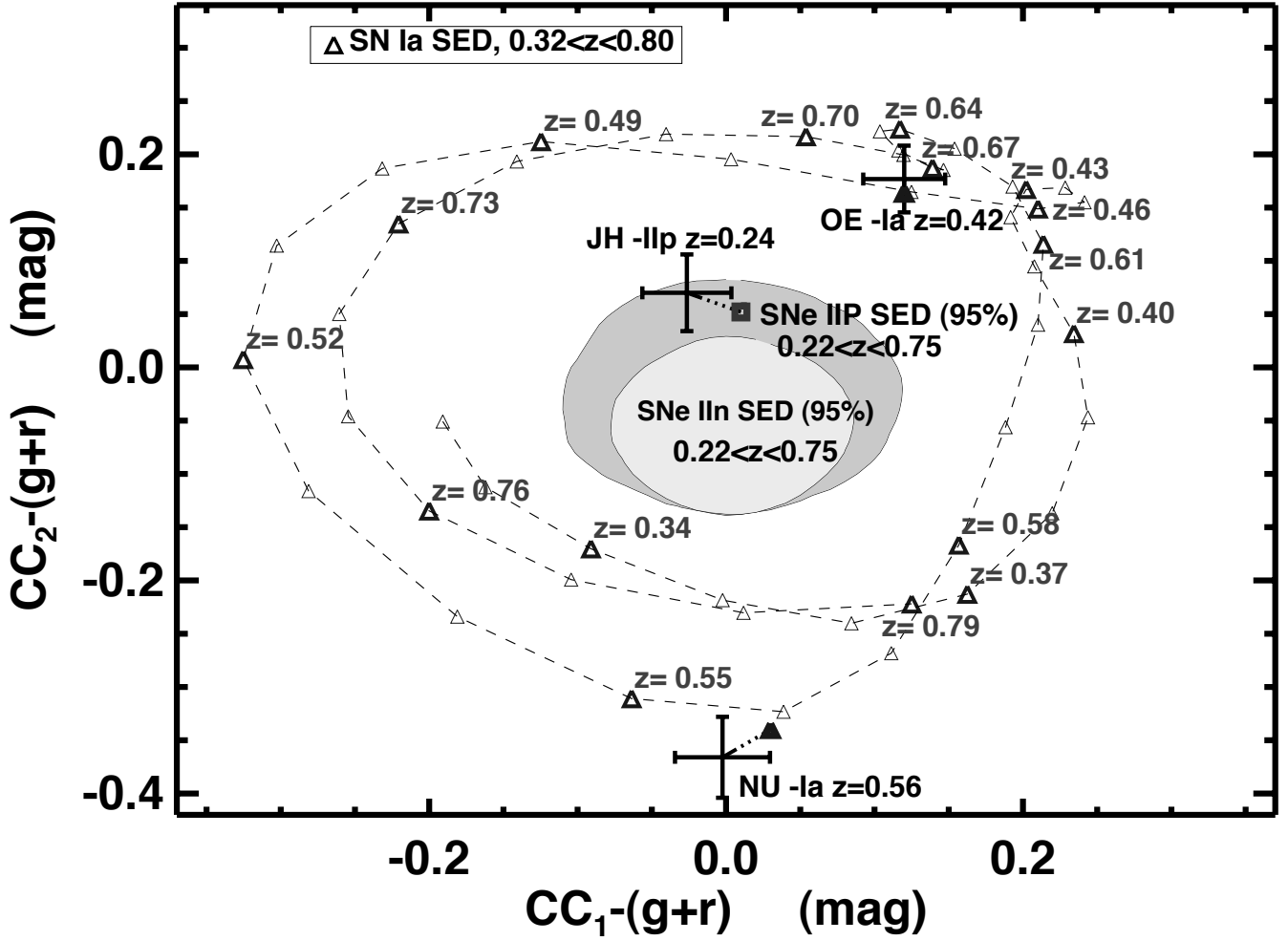}
\end{center}
\caption{
NU, OE and JH in $CC-B$ space (with error bars), as well as their synthetic positions (highlighted symbols).  The synthetic positions of the SNe are within the standard deviations of the photometric positions.  The distribution of SNe Ia in $CC-B$ space appears different than from Fig. \ref{fig:FigCirc3} because here the Megacam sloan g' and r' filters make up the broadband filters.  From the positions of the SNe in $CC-(g+r)$ space, one can find the type of SN, and if Ia, its redshift.}
\label{fig:AllArr}
\end{figure}

\pagebreak
\begin{figure}[h]
\begin{center}
\includegraphics[viewport=1.0in 2.5in 10.0in 10.0in, scale=0.75]{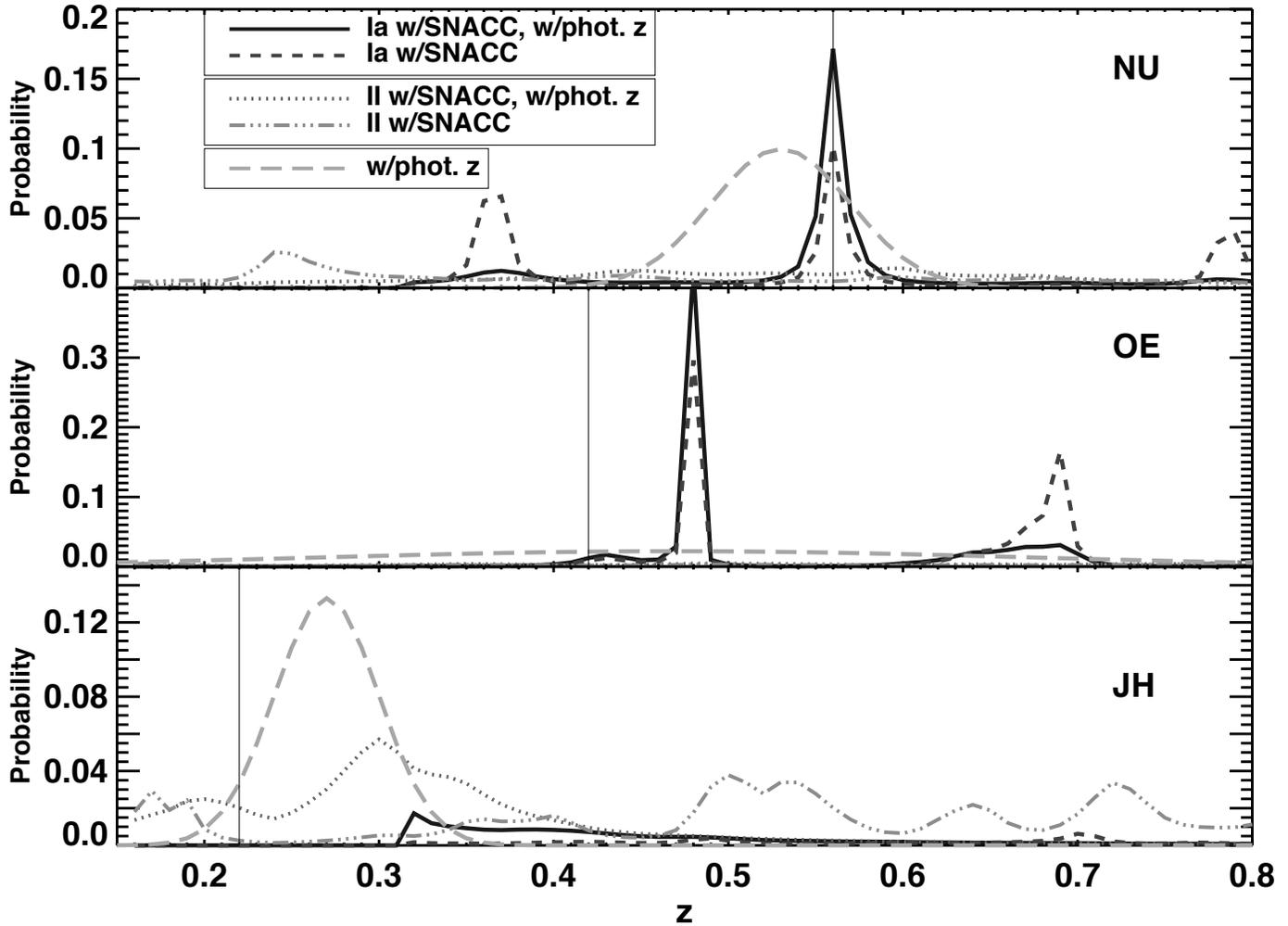}
\end{center}
\caption{
The normalized probability distribution for both redshift and type (SN Ia, SN II) for NU, OE and JH from the chi-squared diagnostic (Eq. 8).  The spectroscopic redshifts are also marked with a solid vertical line at its redshift value.  For NU, the probability distribution for SNe Ia at different redshifts is a gaussian centered at $z=0.56$ with $\sigma_z=0.01$.  The peak is $\sim20\times$ greater than the probability ($<0.01$) of the flat SNe II distribution.  For OE, the SN is clearly Ia, but there are degeneracies in the probability distribution due to the large photometric error and OE's position in $CC-B$ space.  For JH, there is no likely SN Ia candidate.}
\label{fig:FigNUDist}
\end{figure}

\pagebreak
\begin{figure}[h]
\begin{center}
\includegraphics[viewport=1.0in 2.5in 10.0in 10.0in, scale=0.75]{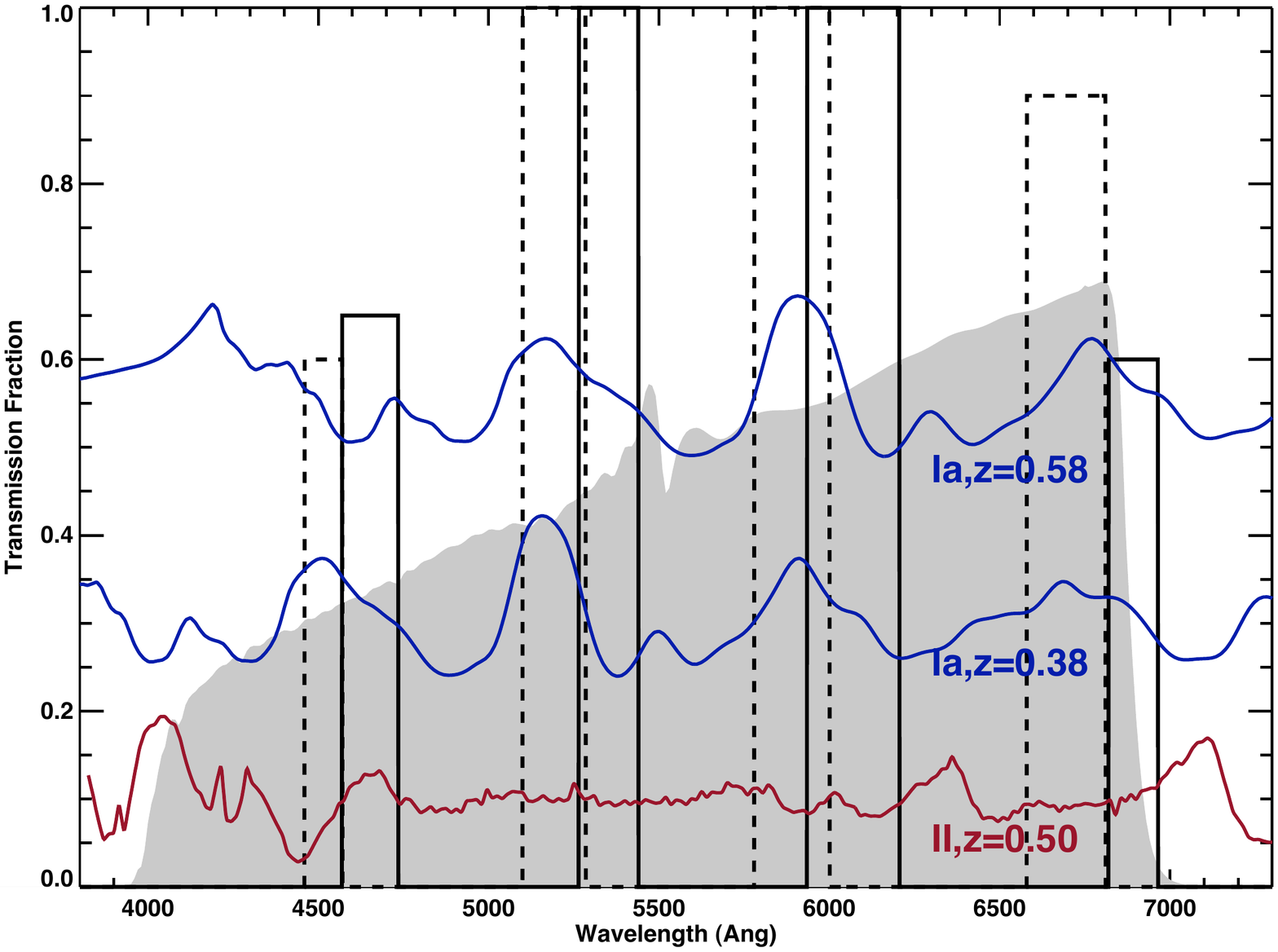}
\end{center}
\caption{
The optimized four-tooth filter design, with one filter portrayed as a solid line, and the other as a dashed line.  The g' and r' SuprimeCam filters are seen in gray. SN Ia spectra at $z=0.38,~0.58$ are overplotted to illustrate how the SNACC filter passbands correlate with the SNe Ia spectral features.  A SN IIP spectrum at $z=0.50$ is also shown to demonstrate little correlation between the SN II features and the filter teeth.  These four-tooth filters have an equivalent width of $\sim600~\AA$.}
 \label{fig:FigFilt4}
\end{figure}

\pagebreak
\begin{figure}[h]
\begin{center}
\includegraphics[viewport=1.0in 2.5in 10.0in 10.0in, scale=0.75]{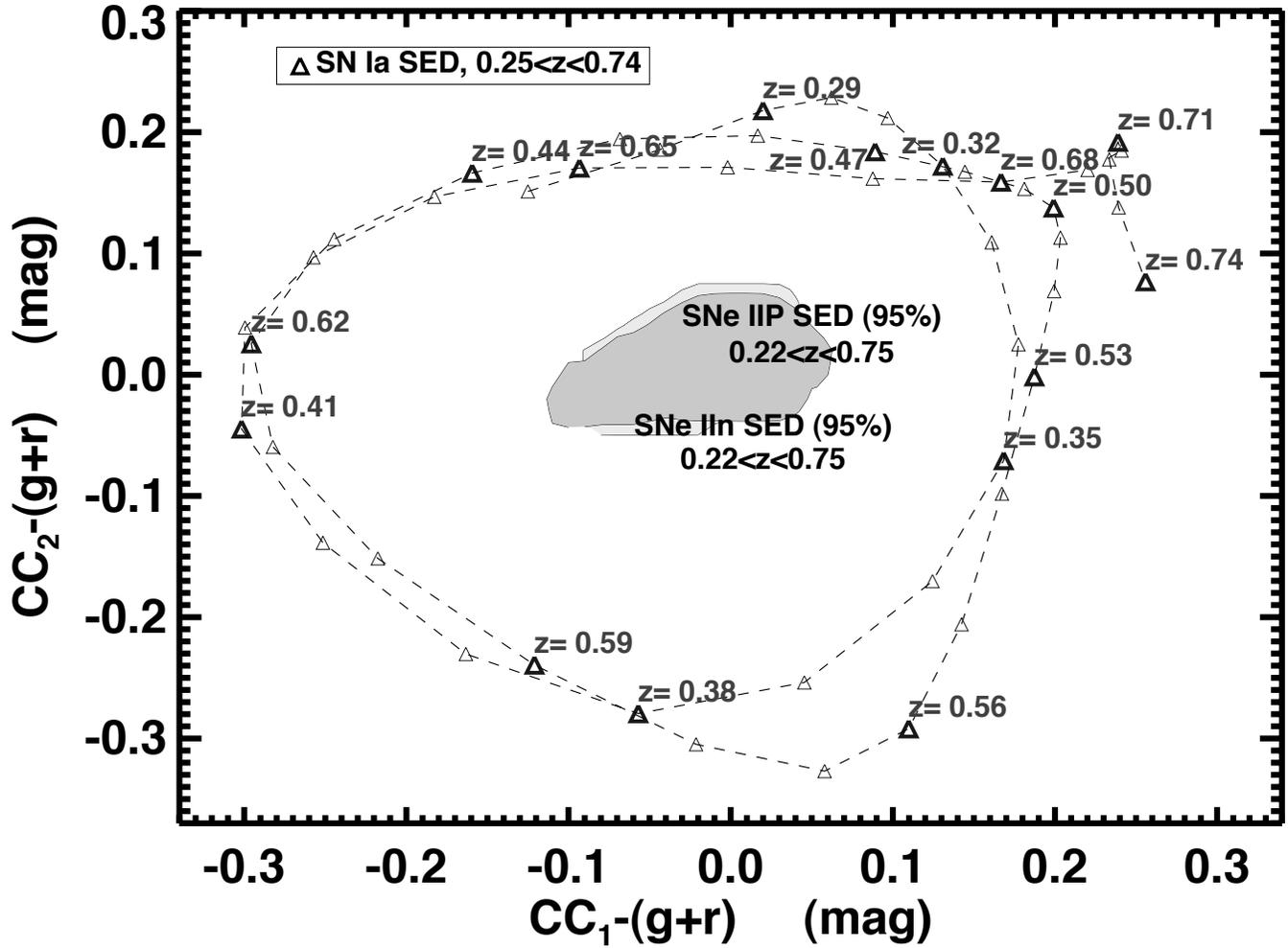}
\end{center}
\caption{
The $CC-B$ diagram for the four-tooth filter design of the SNe Ia (outside) and the SNe II (inside).  While the SNe Ia radius is slightly smaller ($0.20$ mag), the cluster of SNe II is tighter around the center, and there is a similar separation between the SNe Ia and SNe II as from the three-tooth case, except here there is no local degeneracy of the SNe Ia.  The SNe Ia purity and redshift determination for the three-tooth and four-tooth cases are approximately equal.}
 \label{fig:FigCirc4}
\end{figure}
\pagebreak

\end{document}